\documentclass[fleqn,usenatbib]{mnras}

\usepackage{newtxtext,newtxmath}

\usepackage[T1]{fontenc}

\DeclareRobustCommand{\VAN}[3]{#2}
\let\VANthebibliography\thebibliography
\def\thebibliography{\DeclareRobustCommand{\VAN}[3]{##3}\VANthebibliography}

\usepackage{graphicx}	
\usepackage{amsmath}	
\usepackage{threeparttable}

\newcommand{\lya}{Ly$\alpha$\ }
\newcommand{\qsoa}{J1137+3549}
\newcommand{\qsob}{J1602+4228}
\newcommand{\qsoc}{J1630+4012}
\newcommand{\gmodel}{$\Gamma$ model}
\newcommand{\tmodel}{$T$ model}

\newcommand{\rband}{{\it r}}
\newcommand{\iband}{{\it i}}
\newcommand{\nb}{{\it NB816}}

\newcommand{\taueff}{\tau_{\rm eff}}

\newcommand{\mean}[1]{\langle#1\rangle}
\newcommand{\Mean}[1]{\left\langle#1\right\rangle}

\title[the physical origin for scatter of IGM opacity]{The physical origin for spatially large scatter of IGM opacity at the end of reionization: the IGM Ly$\alpha$ opacity-galaxy density relation}

\author[R. Ishimoto et al.]{
\href{https://orcid.org/0000-0002-2134-2902}{Rikako Ishimoto}$^{1}$\thanks{E-mail: ishimoto@astron.s.u-tokyo.ac.jp},
\href{https://orcid.org/0000-0003-3954-4219}{Nobunari Kashikawa}$^{1,2}$,
\href{https://orcid.org/0000-0001-9044-1747}{Daichi Kashino}$^{3}$,
\href{https://orcid.org/0000-0002-9453-0381}{Kei Ito}$^{1,4,5}$,
\href{https://orcid.org/0000-0002-2725-302X}{Yongming Liang}$^{4,5}$,
\newauthor
\href{https://orcid.org/0000-0001-8467-6478}{Zheng Cai}$^{6}$,
\href{https://orcid.org/0000-0002-3800-0554}{Takehiro Yoshioka}$^{1}$,
\href{https://orcid.org/0000-0003-3466-3876}{Katsuya Okoshi}$^{7}$,
\href{https://orcid.org/0000-0002-5464-9943}{Toru Misawa}$^{8}$,
\href{https://orcid.org/0000-0003-2984-6803}{Masafusa Onoue}$^{9,10,11}$,
\newauthor
\href{https://orcid.org/0000-0001-7154-3756}{Yoshihiro Takeda}$^{1}$,
and Hisakazu Uchiyama$^{12}$
\\
$^{1}$Department of Astronomy, Graduate School of Science, The University of Tokyo, 7-3-1 Hongo, Bunkyo, Tokyo 113-0033, Japan\\
$^{2}$Research Center for the Early Universe, The University of Tokyo, 7-3-1 Hongo, Bunkyo-ku, Tokyo 113-0033, Japan\\
$^{3}$Institute for Advanced Research, Nagoya University, Furocho, Chikusa-ku, Nagoya, 464-8601, Japan\\
$^{4}$Department of Astronomical Science, Graduate University for Advanced Studies (SOKENDAI), Mitaka, Tokyo 181-8588, Japan\\
$^{5}$National Astronomical Observatory of Japan, 2-21-1 Osawa, Mitaka, Tokyo 181-8588, Japan\\
$^{6}$Department of Astronomy and Tsinghua Center for Astrophysics, Tsinghua University, Beijing 100084, China\\
$^{7}$ Institute of Arts and Sciences, Tokyo University of Science, 6-3-1, Niijyuku, Katsushika, Tokyo 125-8585, Japan\\
$^{8}$School of General Studies, Shinshu University, 3-1-1, Asahi, Matsumoto City 390-8621, Japan\\
$^{9}$Max-Planck-Institut f\"{u}r Astronomie, K\"{o}nigstuhl 17, D-69117 Heidelberg, Germany\\
$^{10}$Kavli Institute for Astronomy and Astrophysics, Peking University, Beijing 100871, China\\
$^{11}$Kavli Institute for the Physics and Mathematics of the Universe (WPI), University of Tokyo, Kashiwa, Chiba 277-8583, Japan\\
$^{12}$Research Center for Space and Cosmic Evolution, Ehime University, Matsuyama, Ehime 790-8577, Japan.
}

\date{Accepted XXX. Received YYY; in original form ZZZ}

\pubyear{2021}

\begin{document}
\label{firstpage}
\pagerange{\pageref{firstpage}--\pageref{lastpage}}
\maketitle

\begin{abstract}
The large opacity fluctuations in the $z > 5.5$ \lya forest  may indicate inhomogeneous progress of reionization.
To explain the observed large scatter of the effective \lya optical depth ($\taueff$) of the intergalactic medium (IGM), fluctuation of UV background (\gmodel) or the IGM gas temperature (\tmodel) have been proposed, which predict opposite correlations between $\taueff$  and galaxy density.
In order to address which model can explain  the large scatter of $\taueff$, we search for \lya emitters (LAEs) around two (\qsoa\ and \qsob) quasar sightlines with  $\taueff\sim3$  and  \qsoc\ sightline with $\taueff\sim5.5$.
Using a narrowband imaging with Subaru/Hyper Suprime-Cam, we draw LAE density maps to explore their spatial distributions.
Overdensities are found within 20 $h^{-1}$Mpc of the quasar sightlines in the low $\taueff$ regions, while a deficit of LAEs  is found in the high $\taueff$ region.
Although the $\taueff$ of the three quasar sightlines are neither high nor low enough to clearly distinguish the two models, these observed $\taueff$-galaxy density relations all consistently support the \gmodel\ rather than the \tmodel\ in the three fields, along with the previous studies.
The observed overdensities near the low $\taueff$ sightlines may suggest that the relic temperature fluctuation does not affect reionization that much.
Otherwise, these overdensities could be attributed to other factors besides the reionization process, such as the nature of LAEs as poor tracers of underlying large-scale structures.

\end{abstract}




\section{Introduction}
Exploring the evolution of the intergalactic medium (IGM) provides insights into when and how the cosmic reionization proceeded.
The effective \lya optical depth, $\taueff$,  measured in high-$z$ ($z\gtrsim6$) quasar spectra is a useful probe of the IGM state, which is defined as
\begin{equation}\label{eq:tau}
    \taueff = -\ln\left\langle F\right\rangle,
\end{equation}
where $F$ is the observed flux normalized by the intrinsic spectrum.
The observations of  $\taueff$ have been conducted to investigate the IGM state \citep{Fan2006,Becker2015,Eilers2018,Yang2020,Bosman2022}.
The recent measurements of  $\taueff$ revealed a steep increase in $\taueff$ and its scatter at $z>5.5$,
 suggesting a prominent increase in the hydrogen neutral fraction, $f_\ion{H}{i}$, and a spatially patchy reionizing process \citep{Fan2006,Bosman2018,Yang2020,Bosman2022}.
However, a physical origin of such a significant variation in \lya forest opacity at $z > 5.5$ has not yet been identified.

In a photoionized IGM, the ionization equilibrium yields following equation;
\begin{equation}
    n_{\ion{H}{i}}\Gamma = n_{\rm e}n_{\ion{H}{ii}}\alpha_{\ion{H}{i}}(T).
\end{equation}
Here, $\Gamma$ is the photoionization rate, and $n_{\ion{H}{i}}$, $n_{\ion{H}{ii}}$, and $n_{\rm e}$ are the number densities of neutral hydrogen, ionized hydrogen, and free electrons,  respectively.
The $\alpha_{\ion{H}{i}}(T)$ is the radiative recombination coefficient, and $T$ is the gas temperature.
We adopt $\alpha_{\ion{H}{i}}(T)\propto T^{-0.72}$ \citep{Becker2015pasa}.
Therefore, $f_{\ion{H}{i}}$ scales as
\begin{equation}\label{eq:f_HI}
    f_{\ion{H}{i}}\propto n_{\rm H}T^{-0.72}\Gamma^{-1},
\end{equation}
where $n_{\rm H}$ is the total hydrogen density.
\citet{Becker2015} tried to reproduce the observed $\taueff$ by the model where the scatter in $\taueff$ between sightlines is driven entirely by variations in the hydrogen density, $n_\mathrm{H}$ in Equation \ref{eq:f_HI}.
They showed the observed scatter in $\taueff$ at $z<5$ is well reproduced by the simulation; however, at higher redshift, the observed scatter spans wider than that predicted by the simulation.
Their results suggest that the density fluctuation alone is not sufficient to produce the observed large scatter of $\taueff$.

Some possible models have been proposed to explain large fluctuation in $\taueff$.
\citet{Davies2016} built a self-consistent model of the ionizing background that includes fluctuations in the mean free path due to the varying strength of the ionizing background and large-scale density field \citep[\gmodel, see also][]{Daloisio2018,Nasir2020}.
In this model, low-density regions have fewer ionizing sources and a short mean free path, which combine to produce a low ionizing background.
This increases the neutral fraction and hence the \lya opacity.

On the other hand, \citet{Daloisio2015} proposed that large temperature fluctuations may produce the observed large scatter of $\taueff$ (\tmodel).
In this scenario, overdense regions are the most opaque because the gas densities are higher, and also because these regions are reionized first, allowing them to cool.
This means that large $\taueff$ is observed where the galaxy number density is large, contrary to the \gmodel.

Another model is that the $\taueff$ scatter is driven by fluctuations of a radiation field dominated by rare, bright sources such as quasars \citep{Chardin2015,Chardin2017}.
However, this model requires higher number density of quasars than that estimated from observations (\citealt{Kashikawa2015,Onoue2017,Matsuoka2019,Jiang2022}; but see \citealt{Grazian2022}).
The model also suggests earlier \ion{He}{ii} reionization than that expected from observations \citep{Daloisio2017}.

More recently, \citet{Kulkarni2019}, \citet{Keating2020,Keating2020a}, and \citet{Nasir2020} performed radiative transfer simulations  to show that reionization complete at $z\sim5.3$, and 50\% of the volume of the universe is ionized at $z\sim7$.
In their late reionization model, residual neutral gas islands produce the large scatter of \lya optical depth.

Recently, \citet{Davies2018} demonstrated that observations of the galaxy populations in the vicinity of the quasar sightline can distinguish the two plausible competing models, \gmodel\ and \tmodel.
Their simulations predicted that  at the sightline with deep \lya trough, fluctuating ionizing background would show a deficit of galaxies, while, quite the contrary, residual temperature variation would show an overdensity of galaxies. 
It is thus possible to directly distinguish these two predictions through measuring the galaxy distributions at the same redshift.

\citet{Becker2018} conducted a search for \lya Emitters (LAEs) at the sightline of ULAS J0148+0600, which has a giant Gunn-Peterson trough \citep{Gunn1965} spanning 110 $h^{-1}$Mpc, and  an extremely large  \lya optical depth $\taueff\geq7.2$ \citep{Becker2015}.
They found a significant deficit of $z\simeq5.7$ LAEs within 20 $h^{-1}$Mpc from the quasar sightline.
The result is consistent with the prediction with fluctuating UV background model and disfavored the scenario with fluctuating gas temperature.
\citet{Kashino2020} performed a survey of Lyman Break Galaxies (LBGs) at the same ULAS J0148+0600 field.
They also found a deficit of galaxies near the trough, consistent with \citet{Becker2018}, suggesting that the paucity of the LAEs is not purely due to absorption of \lya photons, but reflects a real underdensity of galaxies in this field.
\citet{Christenson2021} analyzed LAE distribution in another high $\taueff$ region, SDSS J1250+3130 with optical depth $\taueff=5.7\pm0.4$, and also found a LAE deficit around the quasar sightline, which are consistent with previous studies.
However, these studies carried out galaxy  searches for only two sightlines. 
Any other interpretations including a genuinely LAE low-density region, cannot be rejected.
Further observations of other quasar sightlines are needed in order to conclude the trend.

In this work, we newly observe three fields around high-$z$ quasar sightlines with both high and low $\taueff$ at $z\sim5.7$. 
Using wide field imaging capability of Hyper Suprime-Cam \citep[HSC;][]{Furusawa2018,Kawanomoto2018,Komiyama2018,Miyazaki2018} mounted on the Subaru telescope, we conduct LAE search at $z\sim5.7$ by using narrow-band filter, NB816 in the fields to measure the spatial distribution of LAEs around three sightlines.
This work includes the targets with low $\taueff$, for the first time, to perform the counter test to contrast what we find for them with the results from high $\taueff$ sample,
investigating which of  the two conflicting models, \gmodel\ and \tmodel\ can explain  the large scatter of $\taueff$.
We compare our results among three fields and the previous studies and discuss the plausible model for the origin of the patchy reionization.

In Section \ref{sec:data}, we present our target selection, observation data, its reduction, and the LAE selection.
We show our results in Section \ref{sec:results}. 
In Section \ref{sec:discussion}, we discuss the implications obtained from the comparison of our observations with the model, after taking into account the uncertainties of the observational data.
Finally, we summarize the paper in Section \ref{sec:summary}.
We assume a $\Lambda$CDM cosmology with $\Omega_m=0.3$, $\Omega_\Lambda=0.7$, and $h=0.7$.
We use the AB magnitude unless specified otherwise.

\section{Observations and sample selection}
\label{sec:data}

\subsection{Target selection} \label{sec:target}

\begin{table*} 
    \centering
    \caption{Overview of the observed fields} \label{tab:field}
    \begin{threeparttable}[htbp]
    
    \begin{tabular}{lcclll}
        \hline\hline
        Field & R.A. & Decl. &quasar redshift&redshift references&$\taueff$\\ \hline
        J1137+3549 & $11^{\rm{h}}37^{\rm{m}}17^{\rm{s}}.73$ &$+35\degr49\arcmin56\arcsec.9$ & $6.009\pm0.010$&\citet{Shen2019}
         & $3.07\pm0.03$ \\
        J1602+4228 & $16^{\rm{h}}02^{\rm{m}}53^{\rm{s}}.98$ & $+42\degr28\arcmin24\arcsec.9$ & $6.083\pm0.005$&\citet{Shen2019}
                   & $3.23\pm0.05$ \\
        J1630+4012 & $16^{\rm{h}}30^{\rm{m}}33^{\rm{s}}.90$ & $+40\degr12\arcmin09\arcsec.7$ & $6.065\pm0.007$&\citet{Carilli2010}
                   & $5.47\pm0.86$ \\
        \hline\hline
    \end{tabular}
    
    Notes - The columns show the field name, coordinate, the quasar redshift and its error, the reference for redshift,  and Ly$\alpha$ optical depth  $\taueff$.
    We measure $\taueff$ within $8080-8274$ \AA, corresponding to $z=5.65-5.81$, 50 $h^{-1}$Mpc range.
    
    \end{threeparttable}
\end{table*}
 \begin{figure*}
    \centering
    \includegraphics[width=\linewidth]{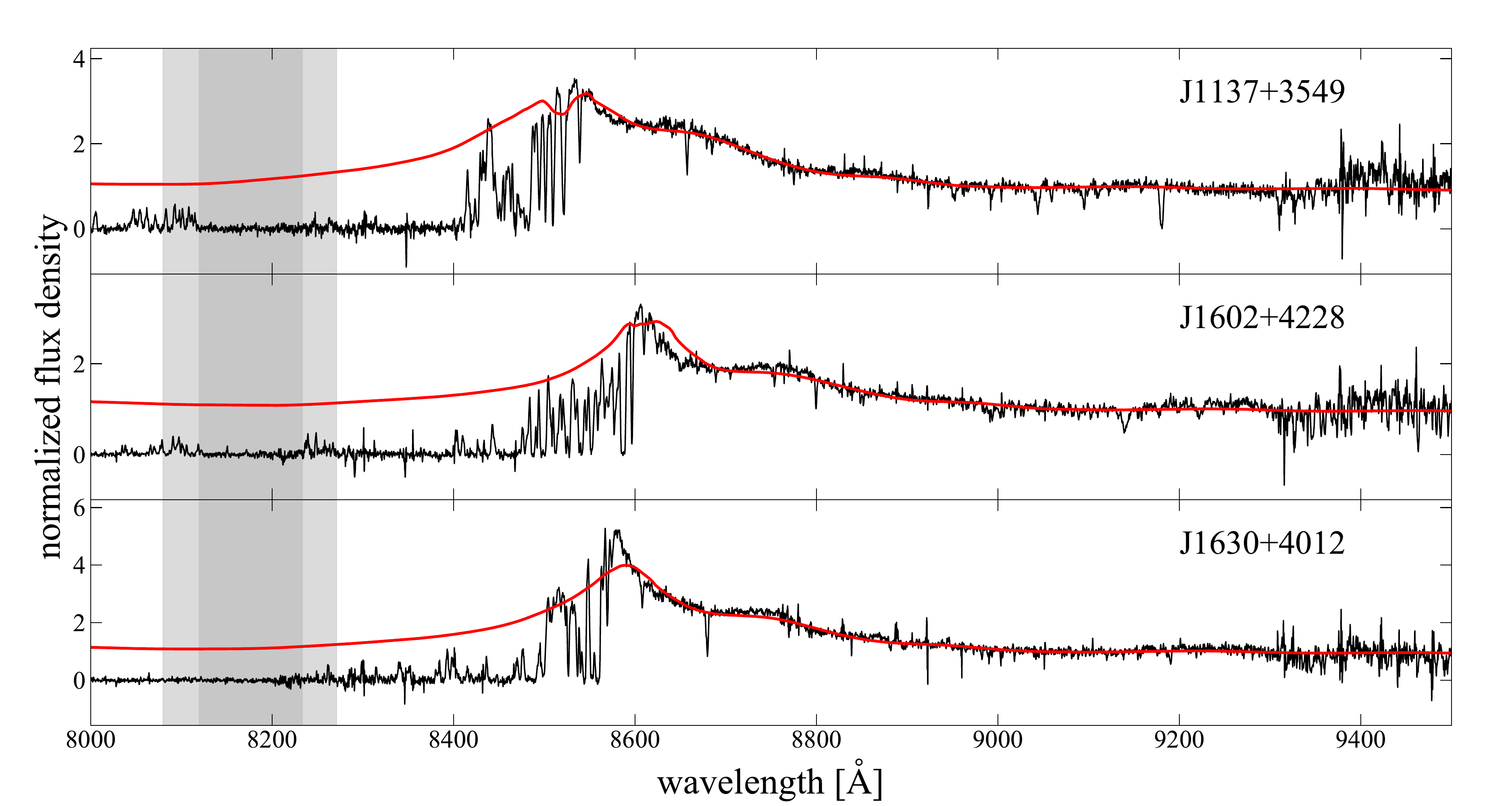}
    \caption{The intrinsic spectra estimated by PCA.
    The black and red curves show the observed and the predicted spectrum of each quasar, respectively.
    The light and dark grey regions indicate the wavelength range corresponding to the FWHM of the NB816 filter and the 50 $h^{-1}$ Mpc range, used for the measurement of $\taueff$.
    The flux density is normalized by the value at 1280 \AA\ in rest frame.
    }
    \label{fig:qspec}
\end{figure*}
We choose three fields to measure the galaxy density at $z\sim5.7$ around sightlines of: \qsoa\ and \qsob\ with low ($\taueff\sim3$) \lya opacities, and \qsoc\ with high ($\taueff\sim5.5$) opacity at $z = 5.726 \pm 0.046$, which corresponds to the HSC NB816 filter's wavelength coverage for \lya emission.
These spectra are selected from the publicly available \texttt{igmspec}\footnote{\url{http://specdb.readthedocs.io/en/latest/igmspec.html}} database \citep{Prochaska2017}.
The details of observed fields are summarized in Table \ref{tab:field}.

There are very few quasars having a high or low \lya opacity just right at $z\sim5.7$ on their spectra; therefore these are almost only solutions among the currently available quasars with good quality and high-enough resolution spectra, though there are certainly some errors in the opacity measurements.
The detail of the $\taueff$ measurement is described in Section \ref{sec:taumeasurement}.
The $\taueff=5.47\pm0.86$ of \qsoc\  exceeds the 95\% range in $\taueff$ predicted by the uniform UV background model from \citet{Becker2015} and cannot be explained by the density variations alone. 
On the other hand, for the low-$\taueff$ region, $\taueff=3.07\pm0.03, 3.23\pm0.05$ are obtained for \qsoa\ and \qsob, respectively.
To clearly distinguish the \gmodel\ and the \tmodel, it is ideal to choose fields with even lower $\taueff$ than those of \qsoa\ and \qsob, although such optimal targets are hard to be found  at $z\sim5.7$.
Based on the $\taueff$ distribution at 5.6<z<5.8 \citep{Bosman2022}, the observed $\taueff=3.07$ (\qsoa) and $3.23$ (\qsob) correspond to 12, 17 percentile from the bottom, while $\taueff=5.47$ (\qsoc) corresponds to 5 percentile from the top.

At first, we had measured $\taueff$ over the FWHM (8120-8234 \AA, 30 $h^{-1}$Mpc) of NB816 filter as an effective wavelength range, and had found the $\taueff=5.58\pm0.62, 6.05\pm0.91, \geq5.06$ for \qsoa, \qsob, and \qsoc, respectively; therefore,
\qsoa\ and \qsob\ were treated as high $\taueff$ regions.
However, when we re-measured more carefully $\taueff$ over 50 $h^{-1}$Mpc range (8080-8274 \AA), which is the same length as that used in the model predictions described later, these two regions have turned out to show low $\taueff$.
Since the spectra of \qsoa\ and \qsob\ show high transmission just outside of FWHM of NB816 as shown in Figure \ref{fig:qspec}, the $\taueff$ values in these fields are greatly sensitive to the width of the $\taueff$ measurement. 
This $\taueff$ uncertainty makes the interpretation of results a little more difficult (see Section \ref{sec:comparison}).
For a fair comparison with the model predictions, we adopt $\taueff$ in 50 $h^{-1}$ Mpc range.
While it may be ideal to choose a sightline target where the transmission is almost constant within the sensitivity range of the NB filter so that it is not affected by the width of the $\taueff$ measurement, it is rare to have extremely long Gunn-Peterson troughs or transparent regions as in J0148+0600. 
Conversely, we would say we are choosing more general sightlines rather than exceptional ones for this study.

\subsection{Optical depth measurement}\label{sec:taumeasurement}

We calculate \lya optical depth, $\taueff$, at the wavelength coverage of  the NB816 filter.
In this part, we present how we estimate intrinsic quasar spectra free from IGM absorption and $\taueff$, following the same manner as \citet{Ishimoto2020}.
The spectra cover the wavelength coverage from 3900 to 10000 \AA\ with a resolution of $R\approx4000$. They are taken from \texttt{igmspec} database \citep{Prochaska2017}.
All three quasar spectra in this study were acquired with the Keck ESI echelle spectrograph.

We estimate the quasar intrinsic spectra after normalizing at rest 1280 \AA\ with principal component spectra (PCS) from a principal component analysis (PCA) of low-$z$ quasar spectra.
 This approach is justified by  the lack of a significant redshift evolution of quasar spectra in the rest-frame UV wavelength range \citep[e.g.,][]{Jiang2009}.
In PCA,  the quasar spectrum, $q_i(\lambda)$, is modeled as a mean quasar spectrum, $\mu(\lambda)$, and  a linear combination of PCS:
\begin{equation}
    q_i(\lambda)\sim\mu(\lambda)+\sum_{j=1}^m c_{ij}\xi_j(\lambda),
\end{equation}
where $i$ refers to a  $i$th quasar, $\xi_j(\lambda)$ is the $j$th PCS, and $c_{ij}$ is the weight.
We use the PCS from \citet{Suzuki2005}, which constructed PCS using low-$z$ ($z<1$) quasars.
First, $c_{ij}'$, the weights for the spectrum redward of 1216\AA, are derived by
\begin{equation}
    c_{ij}' = \int_{1216{\mathrm{\mathring{A}}}}^{\lambda_{\rm upper}}[q_i(\lambda)-\mu(\lambda)]\xi_j(\lambda)d\lambda,
\end{equation}
 where $\lambda_{\rm upper}$ is the upper limit of available wavelength in each observed quasar spectrum.
  \citet{Suzuki2005} produced PCS for 1216\ \AA\ to 1600\ \AA, while our sample has coverages up to $\sim1400$ \AA\ in rest frame.
  
\begin{figure}
    \centering
    \includegraphics[width=1.0\linewidth]{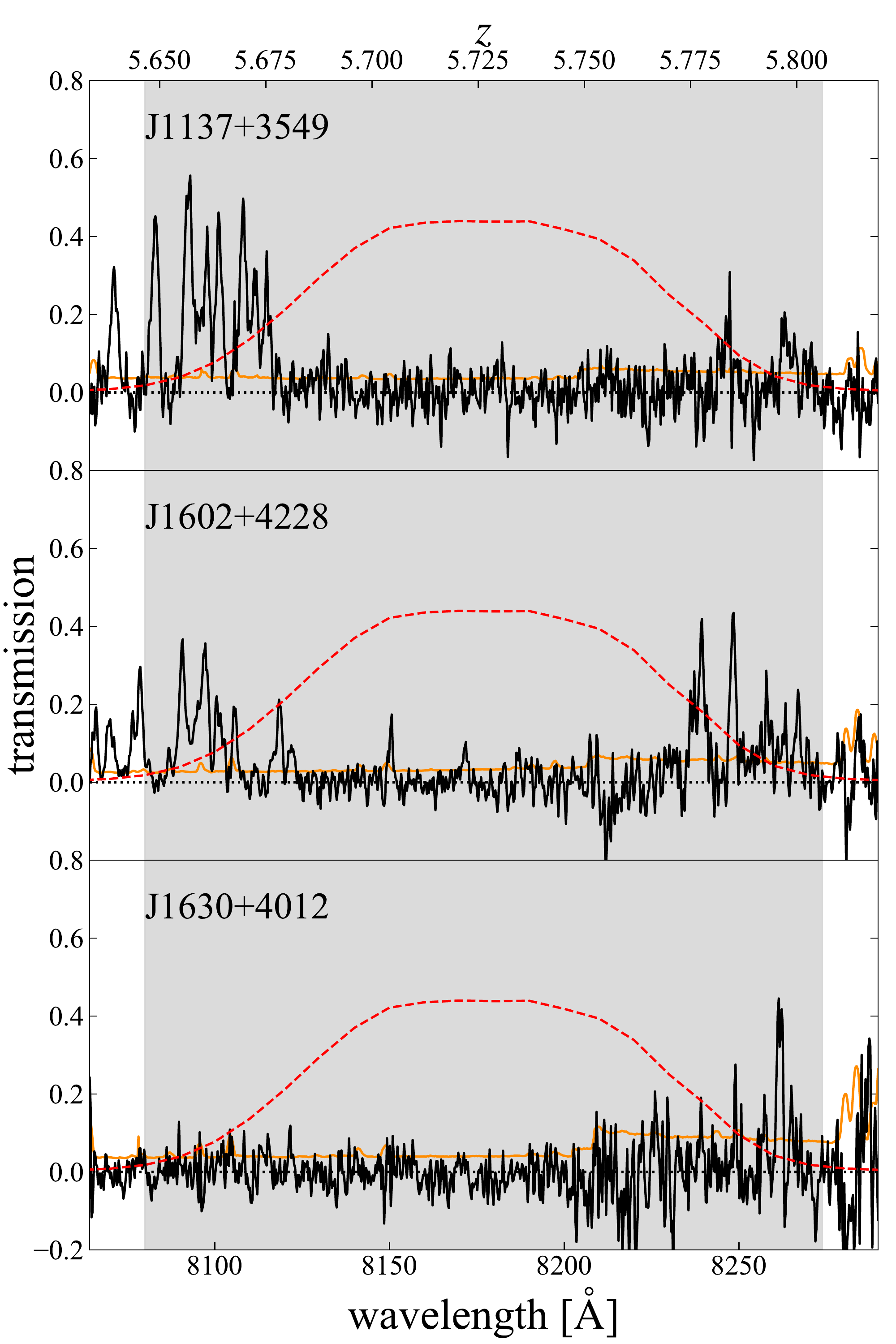}
    \caption{The transmission curves of quasar spectra. The orange lines represent the spectral noise. 
    The grey regions indicate the wavelength range corresponding to the 50 $h^{-1}$ Mpc range, used for the measurement of $\taueff$.
    The red lines present the NB816 transmission curve.
    }
    \label{fig:transmission}
\end{figure}

Then we use the projection matrix $\bmath{X}$ to calculate $c_{ij}$, the weights for the whole intrinsic  spectrum, covering the entire spectral region between 1020\AA\ and 1600\AA, using
\begin{equation}
    c_{ij}= c_{ij}'\cdot\bmath{X}.
\end{equation}
 The projection matrix $\bmath{X}$ is also taken from \citet{Suzuki2005}.
It is the matrix which satisfies the relation $\bmath{C}=\bmath{D}\cdot\bmath{X}$, where $\bmath{C}$ and $\bmath{D}$ are the weights of principal components of the whole and the redward of quasar spectrum derived in \citet{Suzuki2005}, respectively.
We use five PCS for all quasar spectra.
The estimated spectra are shown in Figure \ref{fig:qspec}, and  the continuum-normalized spectrum of each quasar is shown in Figure \ref{fig:transmission}.

We estimate the effective optical depth $\taueff$ using Equation \ref{eq:tau}.
If the mean normalized flux is negative or is detected less than 2$\sigma$ significance, we adopt  a lower limit of optical depth at $-\ln(2\sigma_{\left\langle F\right\rangle})$.
We measure $\taueff$ at $8080-8274$ \AA\ ($z=5.65-5.81$). This range corresponds to 50 $h^{-1}$ Mpc and $>1\%$ transmission of NB816. 

The uncertainty of $\taueff$ is estimated by taking into account the observation error of the spectrum and the uncertainty of the continuum estimation.
Errors due to noise in the spectra are estimated by the Monte Carlo simulation using the noise spectra. 
We generate 100 mock spectra, in which the flux of each spectral pixel is given a random error perturbed within the measured 1$\sigma$ error, and repeat the continuum estimation and $\taueff$ measurement. 
The uncertainty of the $\taueff$ from the continuum error due to the measurement error of the quasar redshift is found to be small, about $0.02$.
However, there are other factors that cause uncertainties in $\taueff$ measurement, which will be verified in Sec \ref{sec:check_tau}.
The measured $\taueff$ and the associated errors are summarized in Table \ref{tab:field}.

\begin{table*}

    \centering
    \caption{Summary of Observations and Imaging Data}\label{tab:datasummary}
    \begin{threeparttable}
    
    \begin{tabular}{lllllc}
        \hline\hline
        Field & Filter & Exp. time [hr]& PSF size  [\arcsec]& $m_{5\sigma}$$^a$&Observation date\\ \hline
        J1137+3549 & HSC-R2 & 0.7 & 0.85 & 26.6 &2020 Feb. 28$^b$\\
                   & HSC-I2 & 1.2 & 0.81 & 26.1 &2019 Apr. 9$^b$, May. 1$^b$, 2020 May. 28$^b$\\
                   & HSC-Z  & 2.0 & 0.77& 25.7&2019 Mar. 16$^b$, 2020 Jun. 22$^b$\\
                   & NB816 & 2.0 & 1.02& 25.4& 2019 Mar. 31\\
        J1602+4228 & HSC-R2 & 2.1 & 0.80& 26.8 &2019 Jun. 11$^b$, 2020 Feb. 23$^b$, 2020 May. 19$^b$\\
                   & HSC-I2 & 1.8 &0.59&26.3&2019 May. 29$^b$, Jun. 8, 9, 2020 May. 20$^b$ \\
                   & HSC-Z  & 2.0 &0.76& 25.8&2019 Mar. 16$^b$, 2020 Jun. 22$^b$\\
                   & NB816 & 1.0 &0.57&25.2&2019 Apr. 9\\
        J1630+4012 & HSC-R2 & 2.1 &1.04&26.6&2019 Jun. 11$^b$, 2020 May. 19$^b$\\
                   & HSC-I2 & 1.5 &0.76&26.2&2019 May. 29$^b$\\
                   & HSC-Z  & 2.0 &0.61& 25.5 &2019 Mar. 16$^b$, 2020 Jun. 22$^b$\\
                   & NB816 & 2.0 &0.62&25.5&2019 Mar. 31, Apr. 9\\
        \hline\hline
    \end{tabular}

     $^a$ the median of $5\sigma$ limiting magnitude of each patch.\\
    $^b$ These data are shared with another program by Kashino et al.(in prep.)
    
    \end{threeparttable}
\end{table*}

\begin{table*}
    
    \centering
    \caption{Summary of the flags for the LAE selection\label{tab:flag}}
    \begin{threeparttable}
    \begin{tabular}{llp{6cm}}
        \hline\hline
        Flag & Value & Comment \\ \hline
        \texttt{detect\_is\_tract\_inner} & True & Object is located closer to the center compared to adjacent tract images\\
        \texttt{detect\_is\_patch\_inner} & True & Object is located closer to the center compared to adjacent patch images\\
        \texttt{base\_PixelFlags\_flag\_edge} & False & Source is outside usable exposure region \\
        \texttt{base\_PixelFlags\_flag\_interpolatedCenter} & False & 	Interpolated pixel in the Source center\\
        \texttt{base\_PixelFlags\_flag\_saturatedCenter} & False & Saturated pixel in the Source center\\
        \texttt{base\_PixelFlags\_flag\_bad} & False & Bad pixel in the Source footprint\\
        \hline\hline
    \end{tabular}
    \end{threeparttable}
\end{table*}

\subsection{Hyper Suprime-Cam Imaging}
\begin{figure}
    \centering
    \includegraphics[width=\linewidth]{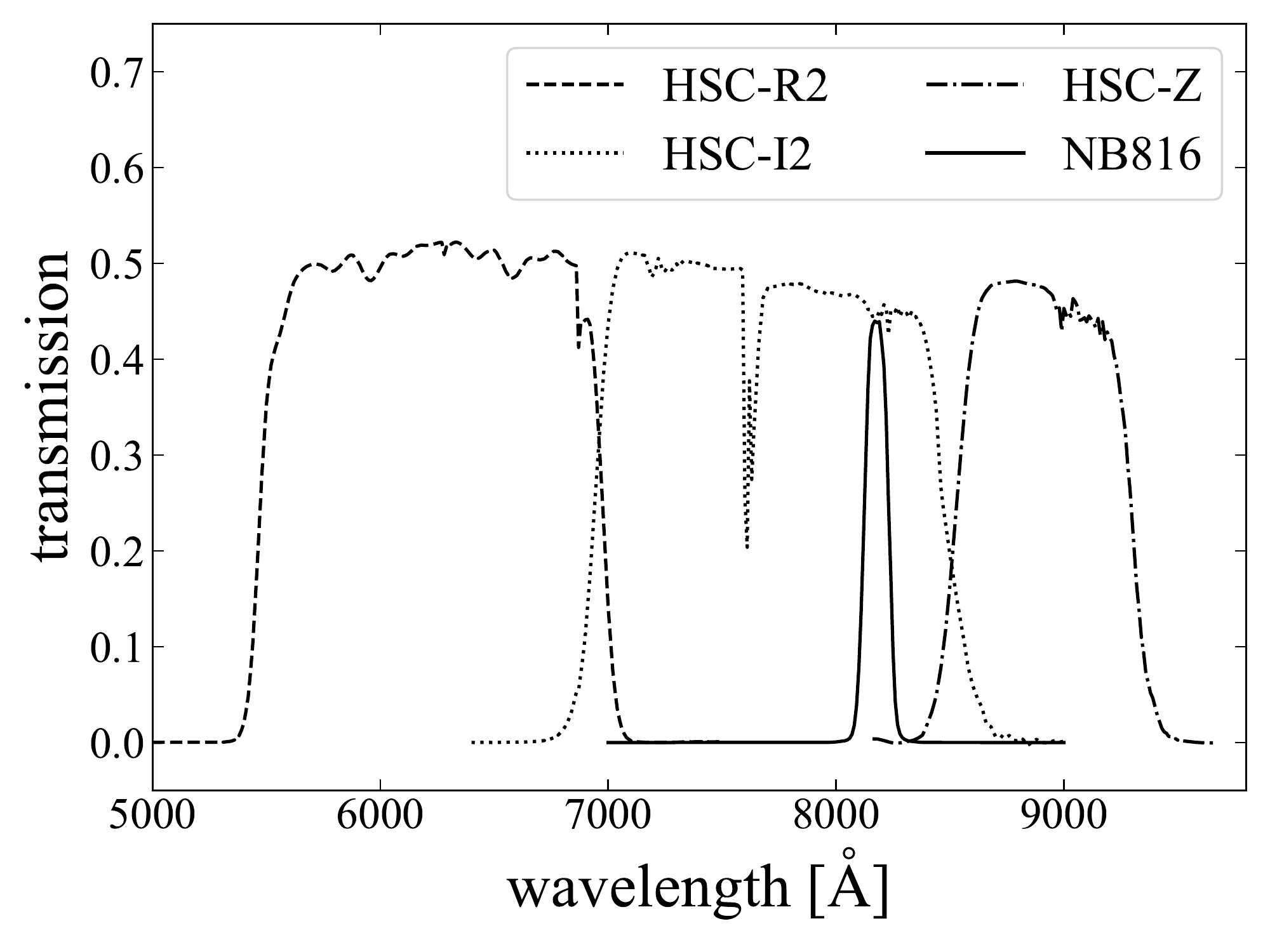}
    \caption{The transmission curves for the filter HSC-R2, HSC-I2, HSC-Z, and NB816.
    Area-averaged transmission data are plotted. }
    \label{fig:filter}
\end{figure}

We use $r$, $i$, $z$, and NB816 imaging data, taken with Subaru HSC, 
 which is a wide-field CCD camera attached to the prime-focus of the Subaru telescope \citep{Miyazaki2018}.
 The HSC has a wide field of view of 1\fdg5 diameter with 116 full-depletion CCDs which have a high sensitivity up to 1 \micron.
The NB816 filter has a transmission-weighted mean wavelength of $\lambda = 8177 $ \AA\  and FWHM = $113$ \AA\ , suitable to detect \lya emission lines  at $z=5.726\pm0.046$.
The filter transmission curves are shown in Figure \ref{fig:filter}.
The $z$-band photometry is not used for the LAE candidate selection, but images are used for the visual inspection.
The observations were performed in 2019-2020 in queue mode.
The single exposure takes $\sim200$ seconds for broad bands, and 600 seconds for NB816.
Exposures were taken at different position angles on the sky around the quasar position to reduce the difference in sensitivity within the field of view as much as possible.
The details of  the observations and the imaging data are summarized in Table \ref{tab:datasummary}.

The data were processed using HSC pipeline version 6.7 \citep[\textit{hscpipe};][]{Bosch2018}.
The \textit{hscpipe} first performs detrending, and calibrates the coordinate and flux using known objects in each shot.
Then it combines images from different exposures and performs detection and photometry of objects.
We apply "forced" photometry, in which photometry is carried out using the centroid coordinate of the reference band for all bands.
The reference band is NB816 in this study.
The total magnitudes and colors are evaluated by  \texttt{convolvedflux\_2\_15}, which corresponds to 1\farcs5 aperture magnitude after an aperture correction.
We use some flags to exclude objects which saturate or are affected by bad pixels.
These flags are summarized in Table \ref{tab:flag}.

\begin{figure*}
    \centering
    \includegraphics[width=1.0\linewidth]{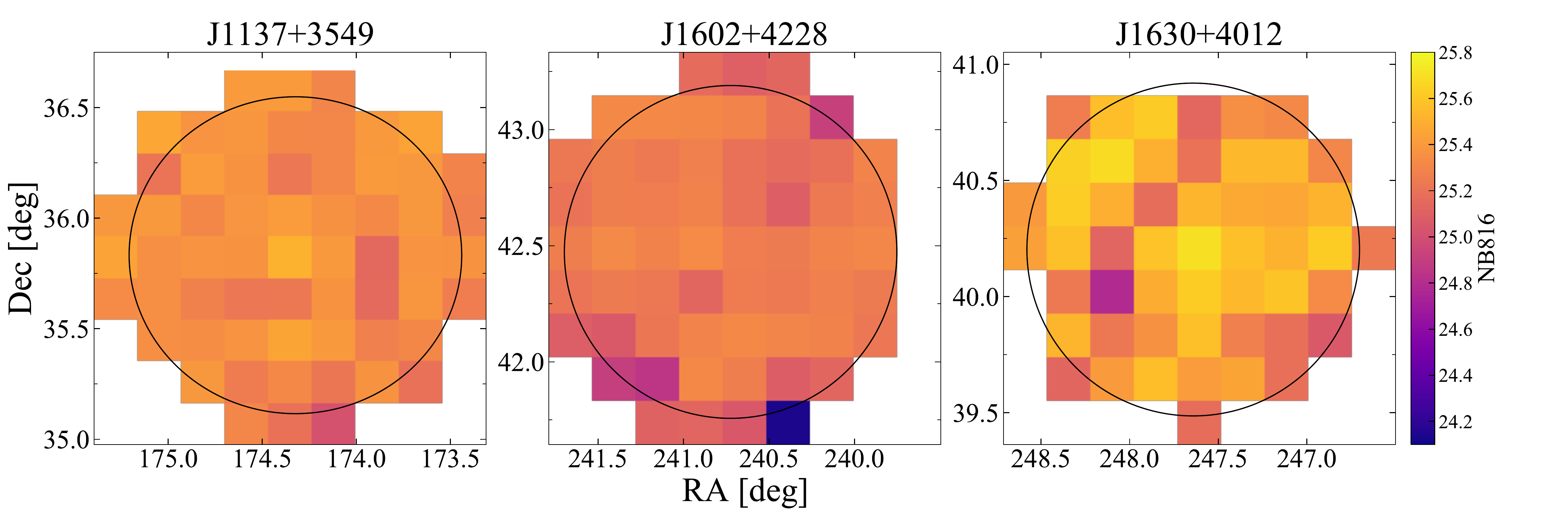}
    \caption{The maps of 5$\sigma$ limiting magnitude of NB816  in the fields of J1137+3549, J1602+4228, and J1630+4012 (from left to right), respectively.
    The black circles indicate the field of view in each field.}
    \label{fig:limmag}
\end{figure*}

We determine the mask regions in addition to the mask defined in \textit{hscpipe} described above.
We build  a star sample brighter than 18 mag in {\it G}-band from the \textit{Gaia} Data Release 2 (DR2) \citep{Gaia2016,Gaia2018}.
We mask regions around the bright stars and exclude the objects which exist in the masks.
The radii of  the mask, $r$, are calculated using the following equation \citep{Coupon2018}:
\begin{equation}
    r\ [\arcsec]= \left\{
    \begin{aligned}
        &708.9 \times \exp(-G_{\rm Gaia})/8.41), & (G_{\rm Gaia}<9) \\
        &694.7 \times \exp(-G_{\rm Gaia})/4.04), & (G_{\rm Gaia}\geq9) 
    \end{aligned}
    \right. ,
\end{equation}
where $G_{\rm gaia}$ is {\it G}-band magnitude from \textit{Gaia} DR2.
We check the final images and set additional masks covering the regions affected by cosmic ray or CCD malfunction.
We calculate the survey area of each field by generating 100 random points per arcmin$^2$ and counting the number of the random points out of the masked regions.
The survey area are 5094, 5025, and 4826 arcmin$^2$ for \qsoa, \qsob, and \qsoc, respectively.

We divide the whole observed area of each field into small square regions, which is called a \textit{patch}, and measure  the limiting magnitude  in each \textit{patch},  following \citet{Inoue2020}.
We conduct 1\farcs5 aperture photometry at 5000 random positions per \textit{patch} avoiding masked regions and the objects detected by \texttt{SExtractor} version 2.8.6 \citep{Bertin1996} in advance. 
The standard deviation, $\sigma$, of the 1\farcs5 aperture photometry was obtained from the histogram of the background-subtracted aperture counts by fitting a Gaussian function. 
The 5$\sigma$ limiting magnitudes of each field are shown in Figure \ref{fig:limmag}.

\subsection{LAE selection}\label{sec:selection}
\begin{figure}
    \centering
    \includegraphics[width=0.9\linewidth]{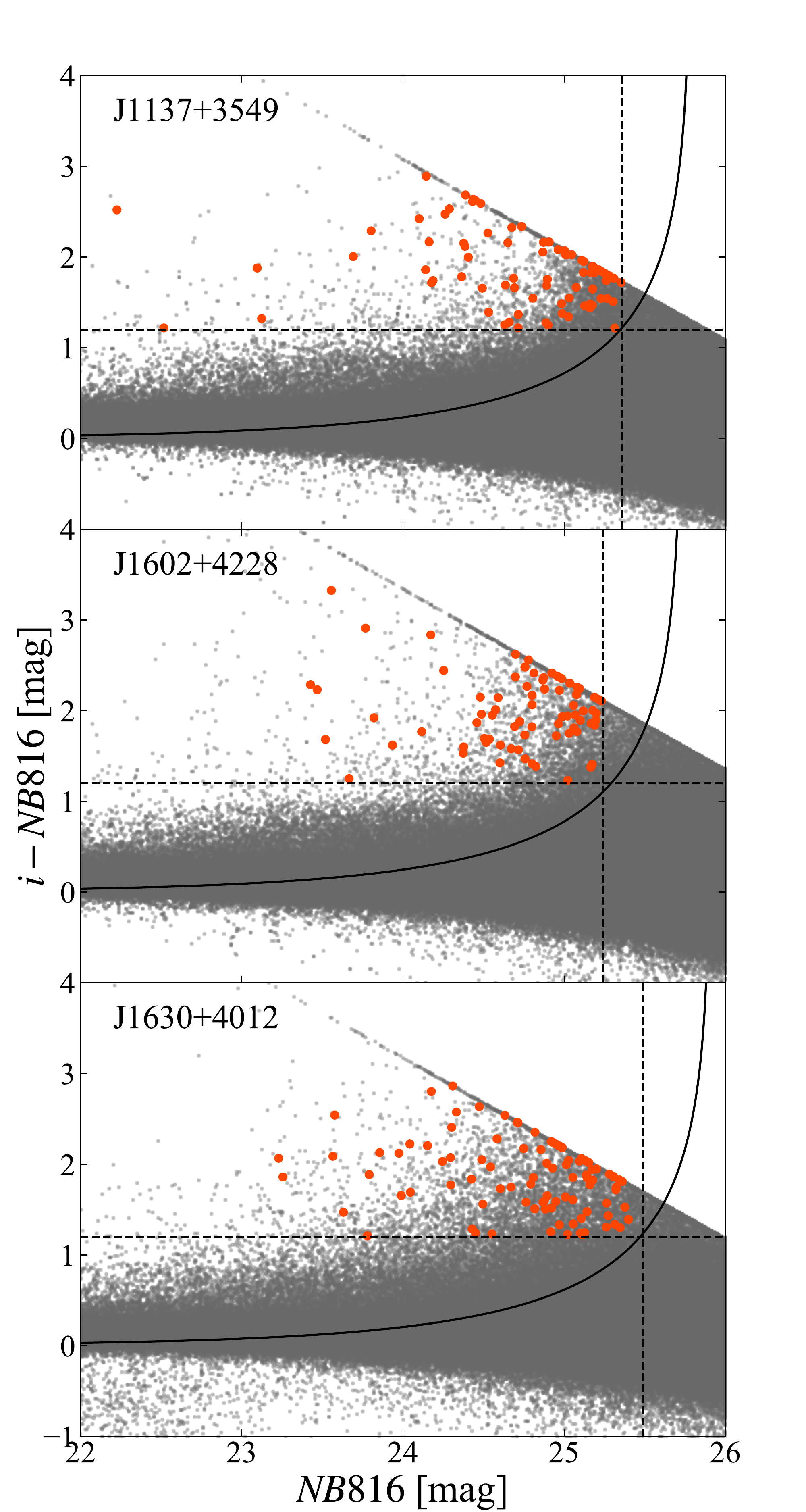}
    \caption{The \nb\ vs. \iband\ - \nb\ diagram of each field.
    The LAE candidates and all detected sources are shown in red and grey dots, respectively.
    The horizontal dashed lines indicate  \iband\ $-$\nb\ $=1.2$ and the vertical dashed lines indicate the median of $\nb_{5\sigma}$ measured in each patch.
    The solid curves indicate Equation \ref{eq:inb>3sigam}.}
    \label{fig:color-mag}
\end{figure}
\begin{figure}
    \centering
    \includegraphics[width=\linewidth]{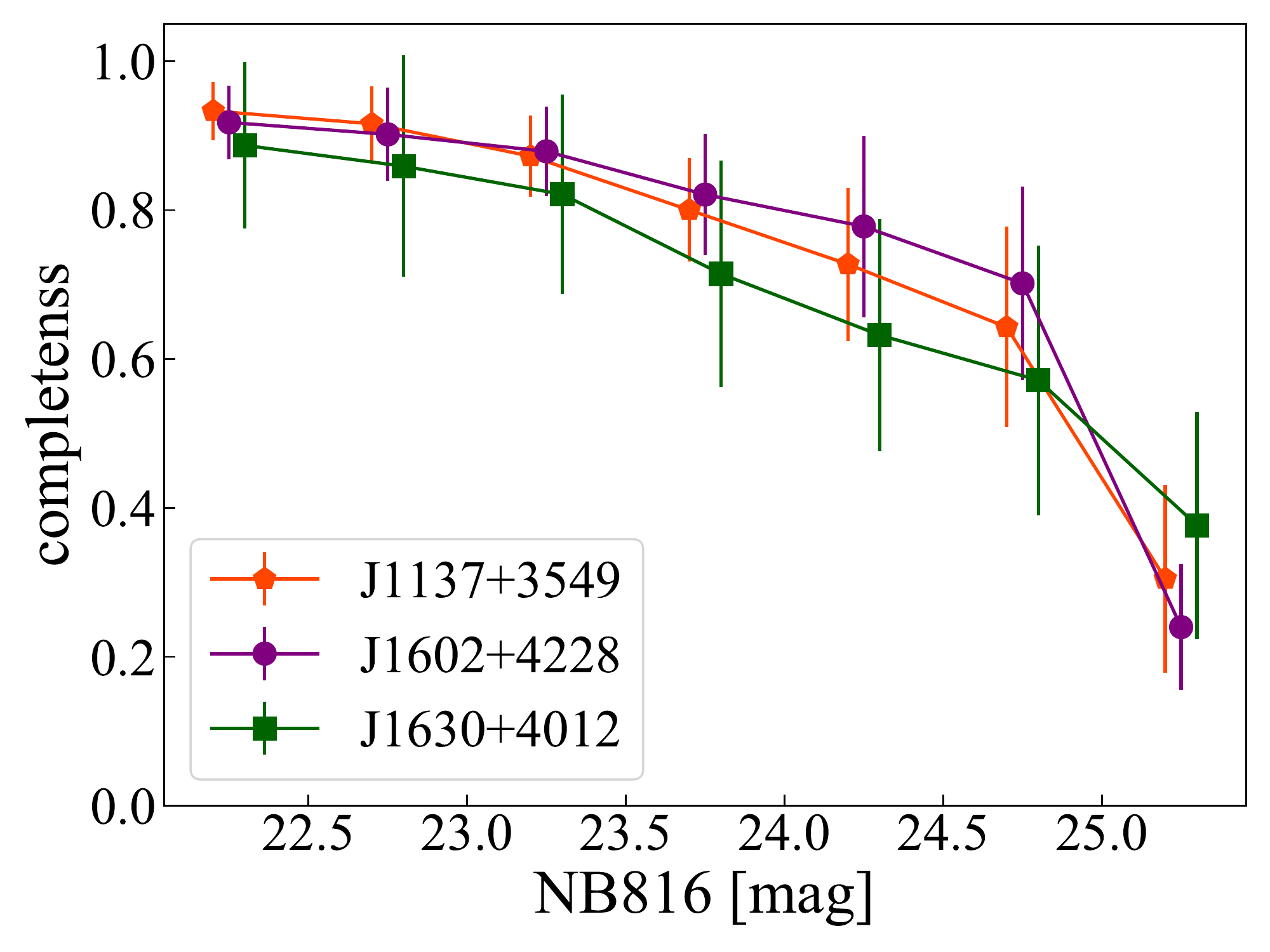}
    \caption{The LAE selection completeness in each field.
    The median values of measured patches are plotted.
    The error bars indicate the Poisson errors.
    The data points are slightly shifted horizontally for clarity.}
    \label{fig:completeness}
\end{figure}
\begin{figure}
    \centering
    \includegraphics[width=\linewidth]{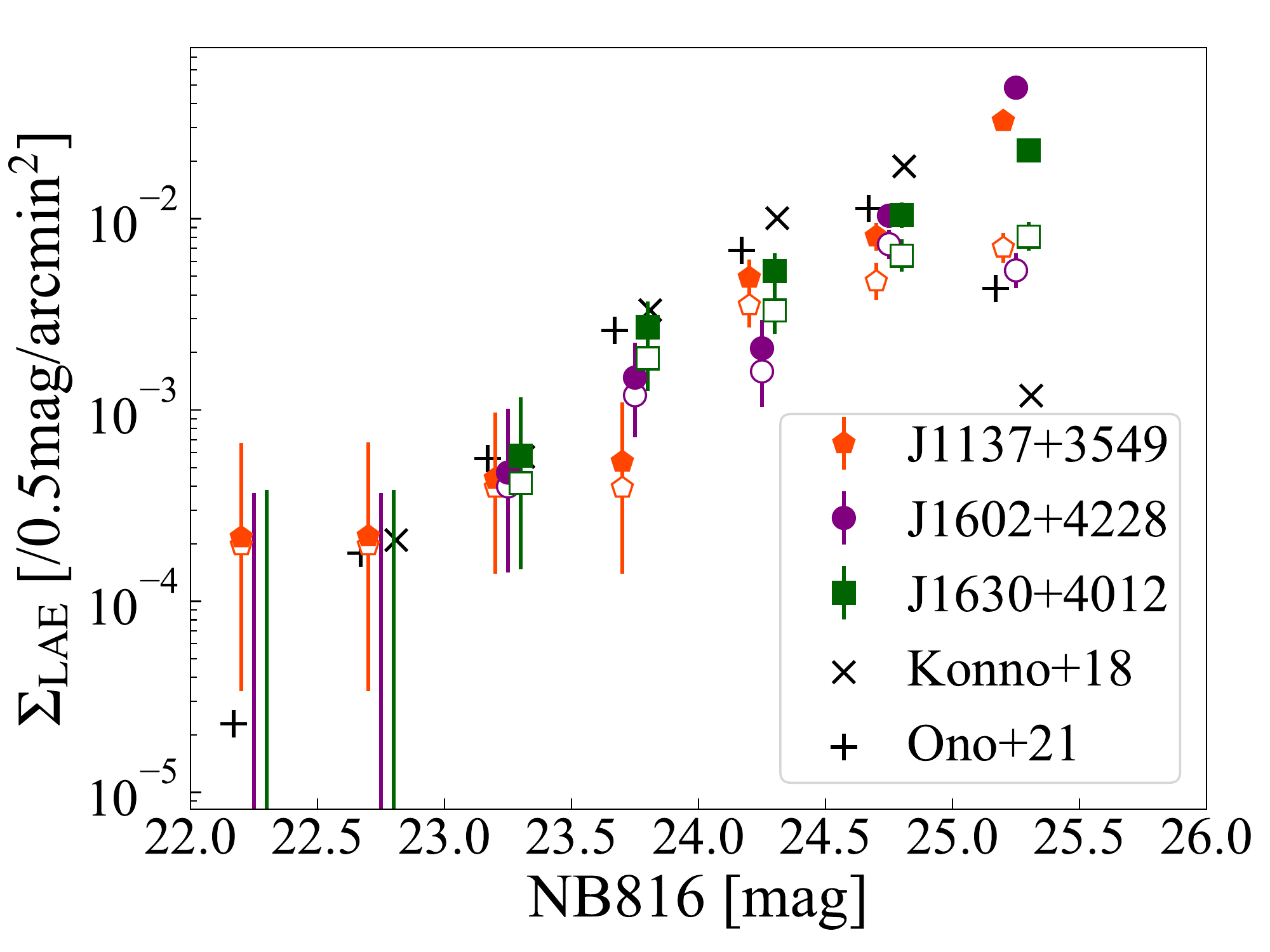}
    \caption{The surface number density of the LAEs with 1$\sigma$ Poisson error in each field.
    The orange pentagons, purple circles, and green squares indicate the surface number densities in the fields of \qsoa, \qsob, and \qsoc, respectively.
    The open and filled symbols show raw and completeness-corrected surface density, respectively.
    The black points show the surface number density from \citet{Konno2018} and \citet{Ono2021}.
    The data points are slightly shifted horizontally for clarity.}
    \label{fig:galcount}
\end{figure}

LAE candidates are selected using the following criteria from \citet{Shibuya2018}:
\begin{align}
    \nb &< \nb_{5\sigma}, \nonumber\\
    \iband - \nb &>1.2,\ {\rm and} \label{eq:i-nb}\\
    \rband > \rband_{3\sigma}\ &{\rm or}\ \rband-\iband>1.0 \nonumber,
\end{align}
where the subscript $n\sigma$ indicates the $n\sigma$ limiting magnitude.
The median values of limiting magnitude measured in each \textit{patch} are used for the selection.
The \iband-band magnitudes fainter than 2$\sigma$ limiting magnitude are replaced with 2$\sigma$ limiting magnitude.
The criterion of $\iband - \nb >1.2$ in Equation \ref{eq:i-nb} corresponds to the rest-frame \lya equivalent width EW$_0 \gtrsim 10$ \AA.
In addition to the criteria described above,
\begin{equation}
    \iband-\nb > 3\sigma_{\iband- \nb} \label{eq:inb>3sigam}
\end{equation}
is also used for the selection.
The $3\sigma_{\iband-\nb}$ is the $3\sigma$ error of $\iband-\nb$ color as a function of the \nb\ flux, given by
\begin{equation}
    3\sigma_{\iband-\nb} = -2.5\log_{10}
    \left(1-3\frac{\sqrt{f_{1\sigma,\nb}^2+f_{1\sigma,\iband}^2}}
    {f_{ \nb}}\right), 
\end{equation}
where $f_{1\sigma,\nb}$ and $f_{1\sigma,\iband}$ are the $1\sigma$ flux error in the \nb\ and \iband\ band photometry, respectively.
Finally, we perform visual inspections for the all LAE candidates selected by the criteria above, in order to reject the objects affected by cosmic ray or bad pixels, noise features in the outskirts of bright objects.
The resultant number of LAE candidates are 84, 80, and 97 for the fields of J1137+3549, J1602+4228, and J1630+4012, respectively. 
Objects from the final LAE candidates are plotted in the \nb\ vs. \iband\ - \nb\ diagram, shown in Figure \ref{fig:color-mag}.

To estimate the completeness of the LAE selection, 
we randomly distribute 100 mock LAEs per 0.5 mag per \textit{patch} in observed images using \texttt{BALROG}\footnote{\url{https://github.com/emhuff/Balrog}}.
We assume a model LAE spectrum with a flat continuum ($f_\nu={\rm const.}$) and a \lya emission included as a $\delta$-function.
The EW$_0$ distribution of \lya emission is distributed to be consistent with that of \citet{Shibuya2018}, and the IGM absorption from \citet{Madau1995} is adopted.
The redshift of LAEs is fixed at $z=5.7$.
We confirmed that the completeness estimates do not significantly change within the observational error when the redshift of the random sources  follow the distribution corresponding to the NB816 transmission curve.
In \texttt{BALROG} workflow, object simulations are performed by \texttt{GALSIM} \citep{Rowe2015}.
The mock LAEs have a S\'{e}rsic index of $n=1.5$ and a half-light radius of $r_c \sim 0.9$ kpc, corresponding to 0.15 arcsec for LAEs at $z=5.7$ \citep{Konno2018}.
The \texttt{BALROG} uses point spread functions calculated by \texttt{PSFEx} \citep{Bertin2011} from observed images.
Then we detect the objects and measure photometry with \textit{hscpipe}, the same way as the detection procedure of observed LAEs.
We apply the same selection criteria of LAEs as described in Equations \ref{eq:i-nb} and \ref{eq:inb>3sigam}.
We performed this procedure for $\sim70$ patches per region.
The medians of measured completeness are shown in Figure \ref{fig:completeness}.

Figure \ref{fig:galcount} shows the surface number density of LAEs.
We corrected for the raw surface number density using the completeness measured in each patch.
The average surface number density after the completeness correction are slightly smaller than those from \citet{Konno2018} and \citet{Ono2021} at $\nb<25$ mag, though the discrepancy is within the range of field-to-field variation, shown by \citet{Ono2021} for five HSC field of views.

\section{results} \label{sec:results}

\begin{figure*}
    \centering
    \includegraphics[width=\linewidth]{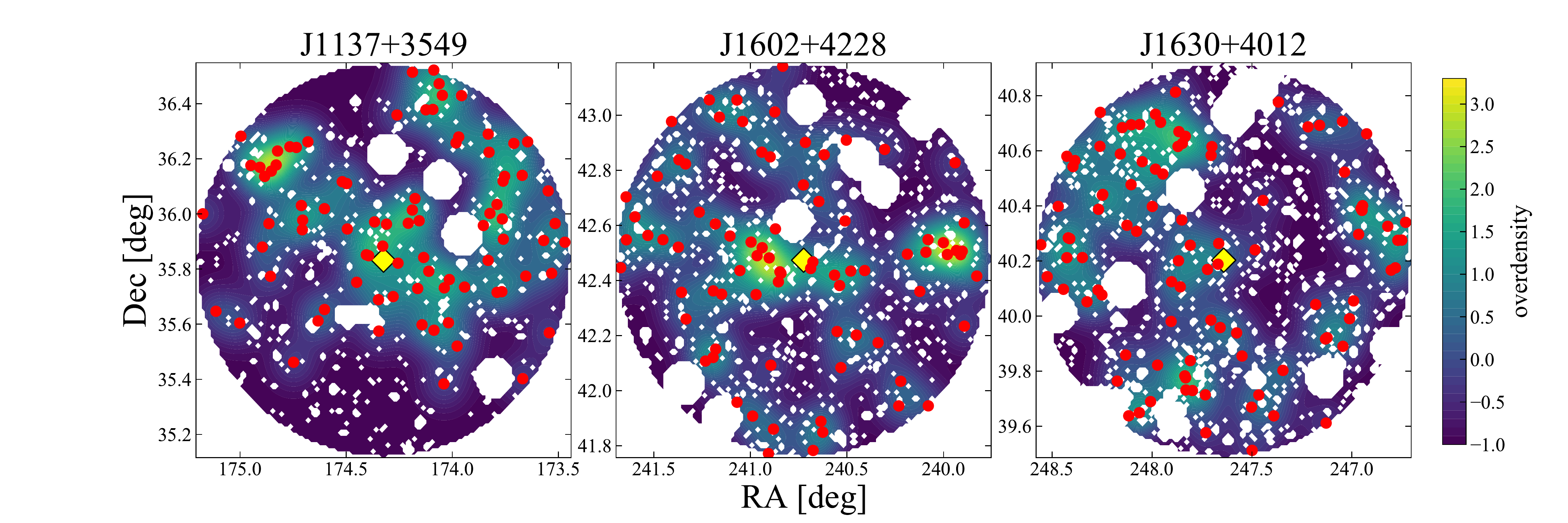}
    \caption{The LAE overdensity maps.  
    Red points show LAE candidates and the yellow diamonds show quasar sightlines.
    The masked regions are shown as the white areas.}
    \label{fig:contour}
\end{figure*}
\begin{figure*}
    \centering
    \includegraphics[width=\linewidth]{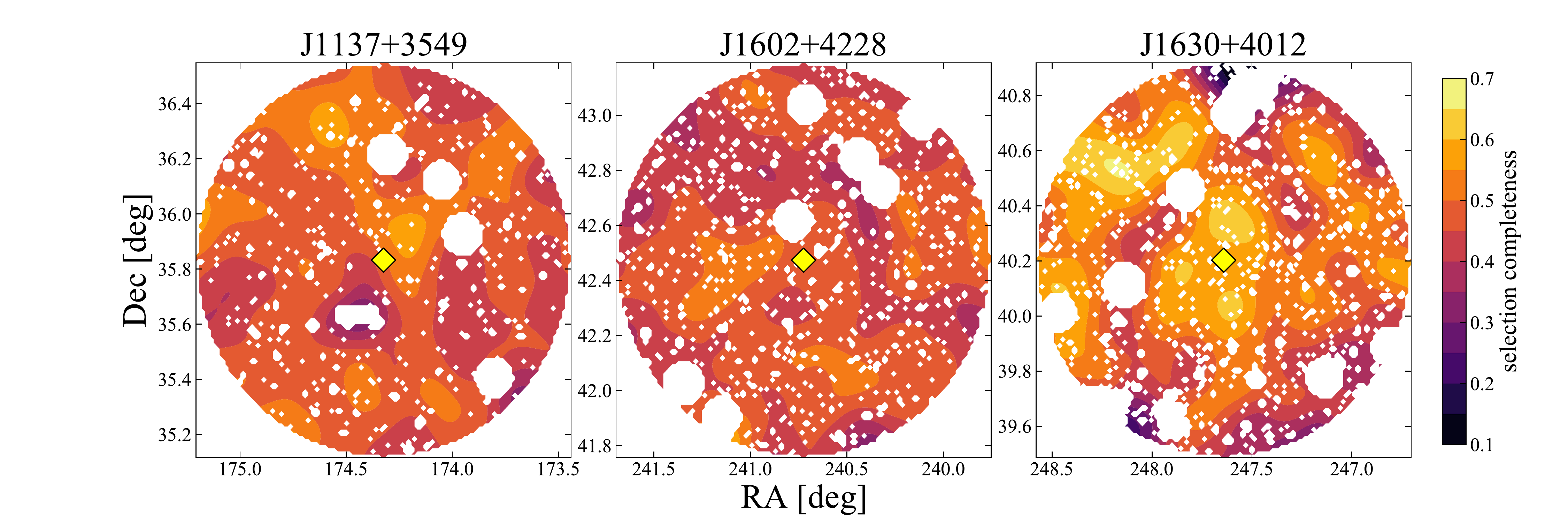}
    \caption{The selection completeness maps.
    The yellow diamonds show quasar sightlines.}
    \label{fig:mockmap}
\end{figure*}

We calculate the  LAE surface number density $\Sigma_{\rm LAE}$ of each field, by summing all LAEs' contribution calculated using 2D Gaussian kernel as $K(\mathbf{X}_i,\mathbf{X}_0)$;
\begin{equation}\label{eq:kernel}
    K(\mathbf{X}_i,\mathbf{X}_0)=\frac{1}{2\pi b^2}\exp\left(-\frac{r(\mathbf{X}_i,\mathbf{X}_0)^2}{2b^2}\right),
\end{equation}
where $b$ is the bandwidth parameter and $r(\mathbf{X}_i,\mathbf{X}_0)$ is the angular distance between two positions.
We apply  a constant bandwidth 4\arcmin.
The maps are constructed through a $128\times128$ grid for each field.
If the area within 8\arcmin\ from a point is covered by a masked region more than 50 \%, we exclude the point from the density map.
The galaxy overdensity $\delta_{\rm LAE}$ is defined as
\begin{equation}
    \delta_{\rm LAE} = \frac{\Sigma_{\rm LAE}-\Mean{\Sigma_{\rm LAE}}}{\mean{\Sigma_{\rm LAE}}},
\end{equation}
where $\mean{\Sigma_{\rm LAE}}$ is the mean surface number density of each field.
The overdensity map (color contour), as well as  the sky distribution of LAEs (red dots) are shown in Figure \ref{fig:contour}.
The measured number densities are corrected using selection completeness maps, which are shown in Figure \ref{fig:mockmap}, taking into account the spatial biases of the sensitivity.
The detail of the estimate of the selection completeness is described in Section \ref{sec:selection}.
We construct completeness maps, weighting the completeness for each magnitude down to the 5$\sigma$ limiting magnitude of \nb. 
The contribution of each mock galaxy is smoothed using Equation \ref{eq:kernel}.

\begin{figure*}
    \centering
    \includegraphics[width=\linewidth]{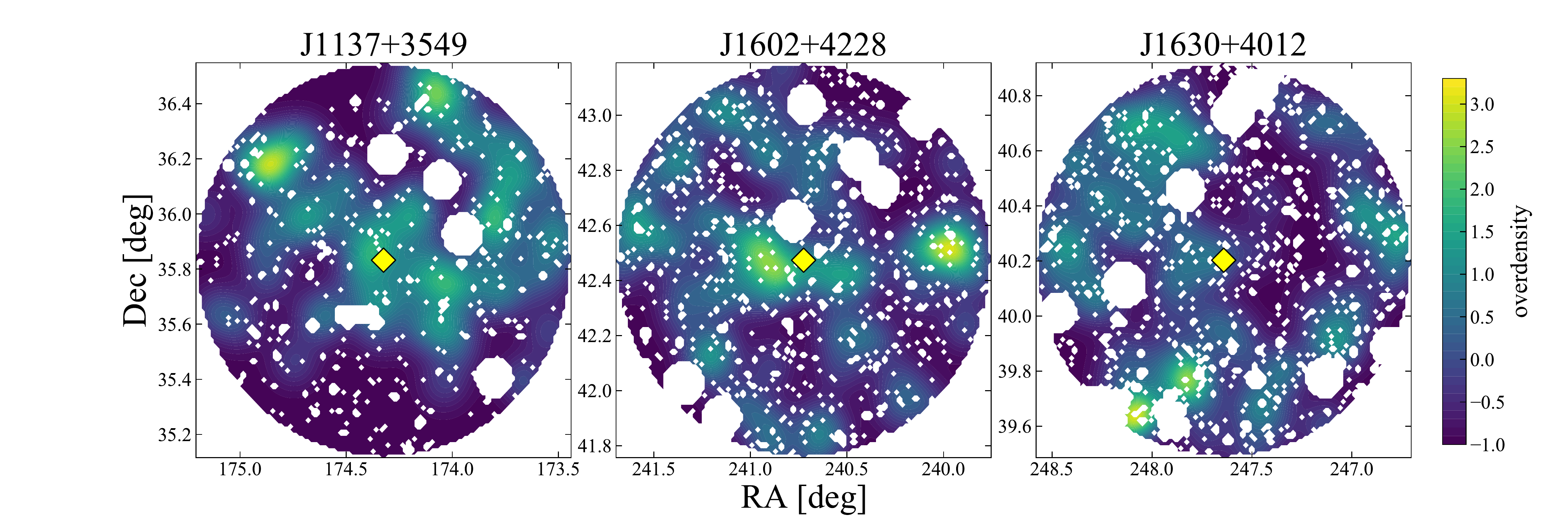}
    \caption{The overdensity maps corrected by the selection completeness maps.
    The yellow diamonds  show the quasar sightlines.}
    \label{fig:odmap}
\end{figure*}

The overdensity maps corrected for the sensitivity variation are shown in Figure \ref{fig:odmap}.
The mean corrected number density  of LAEs brighter than 5$\sigma$ limiting magnitude are  0.018, 0.018, 0.023 arcmin$^{-2}$, for the \qsoa, \qsob, and \qsoc\ field, respectively.
Figure \ref{fig:odmap} shows that there is a large variation of $z=5.7$ LAE densities between the three fields.
The LAE density near the quasar sightline is high in the \qsoa\ and \qsob\ fields, while it is low in the \qsoc\ field.
The overdensities directly above the quasar sightline are $\delta_{\rm LAE}=1.54, 1.28, 0.33$ for the \qsoa, \qsob, and \qsoc\ fields, respectively.

\begin{figure}
    \centering
    \includegraphics[width=\linewidth]{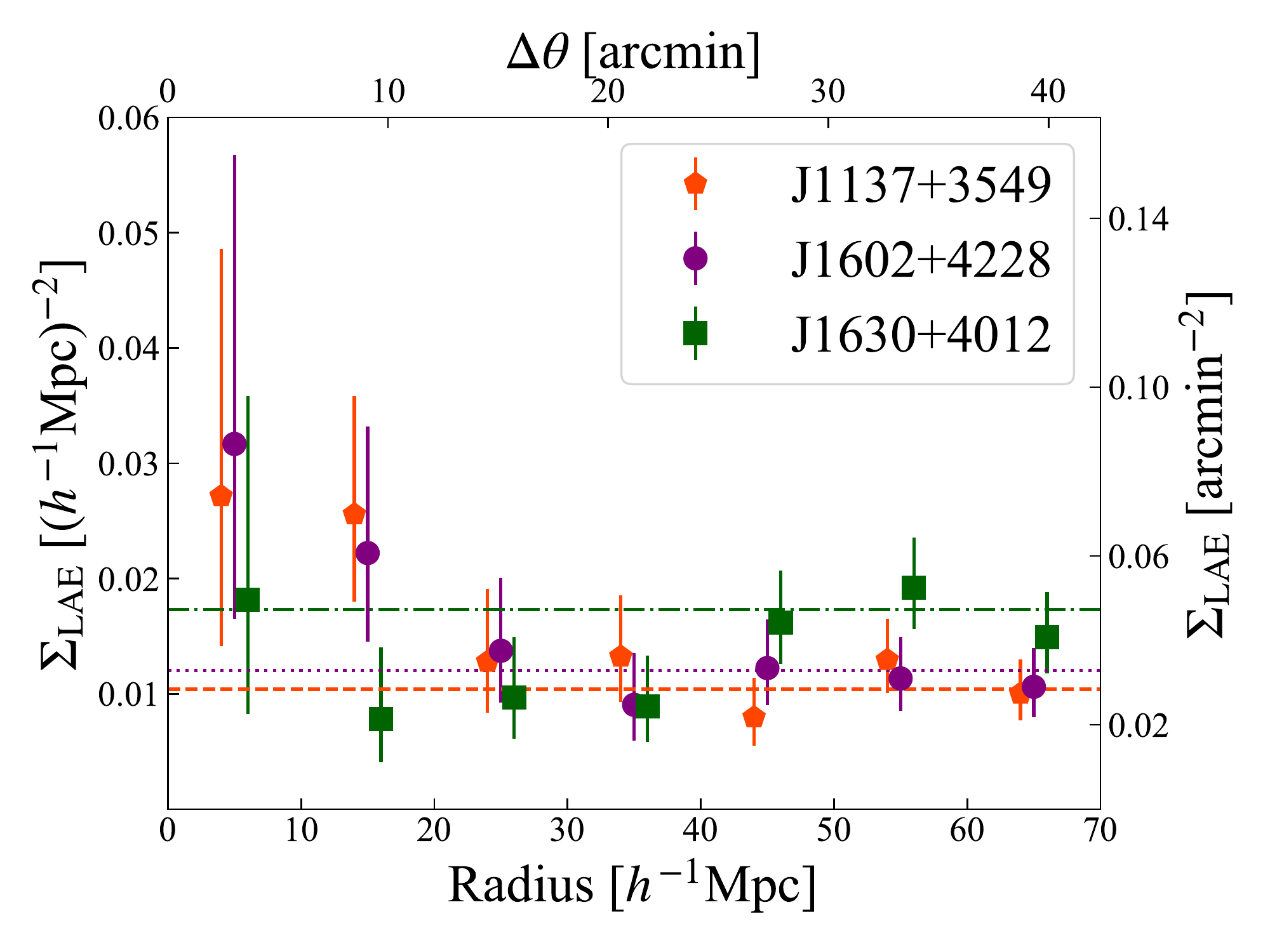}
    \caption{The surface density of LAEs as a function of projected distances from the quasar sightline.
    The orange pentagons, purple circles, and green squares indicate the surface number densities with 1$\sigma$ Poisson errors in the fields of \qsoa, \qsob, and \qsoc, respectively.
    The horizontal dashed, dotted, and dashed-dotted lines show the mean surface density at $>40 h^{-1}$Mpc in the fields of \qsoa, \qsob, and \qsoc, respectively. 
    The data points are slightly shifted horizontally for clarity.}
    \label{fig:radial_density}
\end{figure}

Figure \ref{fig:radial_density} shows the surface density of LAEs as a function of
projected distances from the quasar sightline.
The profiles for the \qsoa\ and \qsob\ fields show a remarkable tendency to rise toward the center,
while that for the \qsoc\ field shows a trend that decreases toward the center except for the innermost bin.
The 5$\sigma$ limiting magnitude of NB816 varies up to $\sim$0.3 mag among the three fields. 
Fixing the limiting magnitude does not change the overall trend, as shown in Appendix \ref{sec:appendix}.
The number densities are consistent with each other at $>20\ h^{-1}$Mpc bins in the  \qsoa\ and \qsob\ fields, and at $>40\ h^{-1}$Mpc in the \qsoc\ field.
 The statistical significance of over(under) densities calculated within 20 $h^{-1}$Mpc from quasar sightline are $2.3\sigma$, $2.7\sigma$, and $1.3\sigma$, respectively, where $\sigma$ is the rms of the fluctuation in the field of view of the number density within an aperture of the same size. 
In the \qsoc\ field, the overdensity directly above the sightline is close to the mean, but when the overdensity is calculated in $<40$ $h^{-1}$Mpc, it yields $\delta_{\rm LAE}=-0.26$, suggesting the LAE deficit.

\section{discussion} 
\label{sec:discussion}
\subsection{Uncertainties of the optical depth measurement}\label{sec:check_tau}

We have described the $\taueff$ measurement in Section \ref{sec:taumeasurement}, but due to the noisy spectra and the difficulty in calibration of high-resolution spectra, the uncertainties of $\taueff$ may not be small.
In this section, we assess the robustness of the $\taueff$ measurement by several methods.

First, We estimate upper limits on $\taueff$, using only the flux of prominent peak transmission, since the overall flux must be equal to or greater than flux in the peaks.
The flux in 8080-8140, 8176-8183, 8243-8248, and 8261-8271 \AA\ for \qsoa, 8086-8108, 8115-8125, 8148-8152, 8168-8174, 8235-8242, and 8247-8274 \AA\ for \qsob, and 8115-8123, 8259-8263 \AA\ for \qsoc\ are used, and for the rest in NB816 coverage, the flux is replaced with zero.
We confirm these wavelength ranges are free from sky OH emissions \citep{osterbrock1996}.
We find the upper limits of $\taueff\leq3.04$ for \qsoa, $\leq3.21$ for \qsob, and $\leq5.30$ for \qsoc.

Another way to check the $\taueff$ measurement is using the NB816 and z-band 1\farcs5 aperture photometry at the quasar sightline.
We use the z-band photometry to estimate the unabsorbed continuum flux at NB816 wavelength by using the intrinsic spectra estimated by PCA to be compared with the observed NB816 flux.
This calculation finds $\taueff = 3.61\pm0.03$, and $5.48\pm0.40$ for \qsoa\ and \qsoc, respectively.
This method could not be applied to \qsob\ because there is an object with comparable-brightness near the quasar sightline, which blended with the quasar image in the z-band, and the accurate continuum flux can not be measured.

Finally, as mentioned in Section \ref{sec:taumeasurement}, $\taueff$ measurements do not rely much on continuum estimation, but for Keck ESI echelle data, it is very difficult to calibrate the zero point correctly, especially for noisy spectra.
To investigate the effect of uncertainty of the zero point of quasar spectra, we calculate $\taueff$ with changing zero point by 1\% of the continuum flux.
Increasing (decreasing) zero point gives us $\taueff=3.31\pm0.05~(2.87\pm0.3)$ for \qsoa, $3.52\pm0.05~(3.01\pm0.03)$ for \qsob, and $\geq5.44~(4.25\pm0.18)$ for \qsoc.
Note that this zero point uncertainty can independently add to the uncertainties of the above two $\taueff$ evaluations.

In summary, each measurement agrees almost consistently with the others: the $\taueff$ of J1137+3549 and J1602+4228 are
$\sim3$, while the $\taueff$ of J1630+4012 is $\taueff\sim5.5$, although there are non-negligible uncertainties.

\subsection{Comparison with the models}\label{sec:comparison}

\begin{figure}
    \centering
    \includegraphics[width=\linewidth]{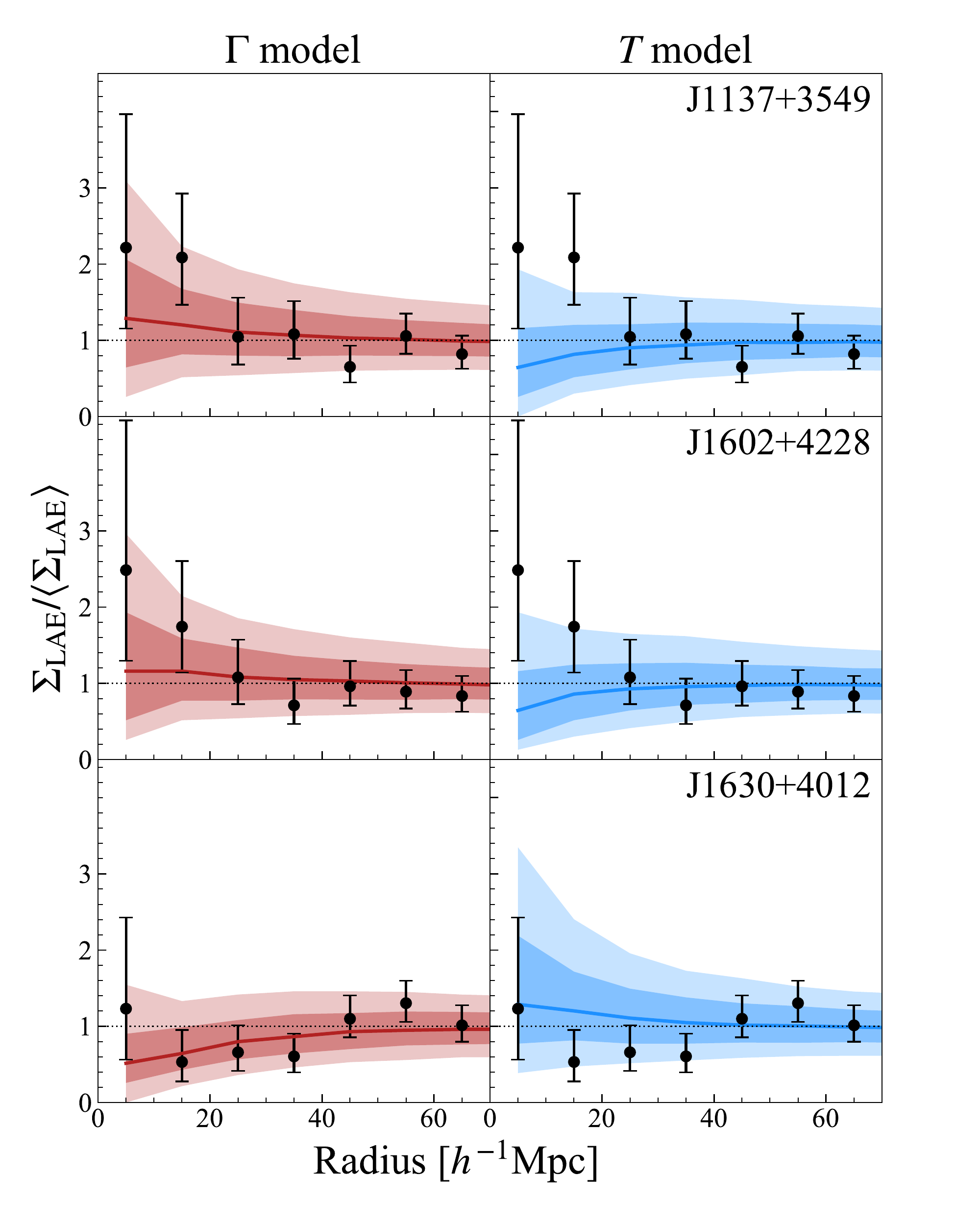}
    \caption{The surface density of LAE candidates normalized by the mean density in each field, as a function of projected distances from the quasar sightline.
    The error bars indicate 1$\sigma$ Poisson errors.
    The red and blue lines show the mean model predictions for the \gmodel\ and the \tmodel\ for fields surrounding 50 $h^{-1}$Mpc sightline for each observed central $\taueff$.
    The dark and light shaded regions indicate the 68\% and 95\% ranges from random trials from the model predictions.}
    \label{fig:radial_los}
\end{figure}

A deficit of galaxies is predicted at the sightline with deep \lya trough based on the \gmodel, while in quite contrary, an overdensity of galaxies is predicted based on the \tmodel.
This study found that LAE underdensity around a high $\taueff$ region and overdensities around low $\taueff$ regions at $z=5.7$, being qualitatively consistent with the \gmodel; however, a more quantitative comparison is needed. 
In this section, we compare our observations with various reionization models.

\subsubsection{\gmodel\ and \tmodel}\label{sec:both_model}
We compare the observational results with the model predictions of \citet{Davies2018}, which provides model predictions of their radial distribution of $\Sigma_{\rm LAE}$ for various central $\taueff$ for the $\Gamma$ and $T$ models.
The model that we used here is almost identical to the one in their original paper, but in order to match our observations as closely as possible, we asked them to change the limiting magnitude from their original of $\nb=26.0$ to $\nb=25.5$, which is almost the same as the limiting magnitude of the observed data, $\nb=25.2-25.5$ mag.
We note that the detection completeness in the model is assumed to be 100\% down to this limit, which makes the uncertainty of the predictions smaller than that of the actual observation.
This model assumes the IGM $\taueff$ measurement over sightline 50 $h^{-1}$Mpc, corresponding to $z=5.65-5.81$.
However, it should be noted that the observed LAE density is weighted by the transmission rate of the NB816. Therefore, what should be
exactly compared to this LAE density is the model prediction based on $\taueff$ weighted by NB816 transmission rate, which is not exactly the same as their 50 $h^{-1}$Mpc model.

Figure \ref{fig:radial_los} shows the surface density of LAEs normalized by the mean density in each field as a function of projected distances from the quasar sightline, along with the model predictions based on the \gmodel\ \citep{Davies2016} and the \tmodel\ \citep{Daloisio2015}.
The model predictions assume LAEs down to $\nb = 25.5$ and the observed $\taueff$ measured at each quasar sightline.
The profiles for the \qsoa\ and \qsob\ fields are rather consistent with the \gmodel. 
However, the model predictions have large uncertainties and are too similar to distinguish at the $\taueff$ of \qsoa\ and \qsob. 
All the data points are consistent with both model predictions within the errors.
The central $\taueff$ here is simply measured in the range of 50 $h^{-1}$Mpc for both the observation and the model; therefore, a more detailed $\taueff$ measurement that takes NB transmissions into account, as was done for LAE selection, might be a more effective way of testing these models.
The profile for the \qsoc\ field, whose overall trend is to decrease toward the center, is consistent with the prediction by the $\Gamma$ model.
The excessively large relative density in the central region in the two low-$\taueff$ fields may be due to either the intrinsic large-scale structure (see Sec. \ref{sec:laebias}) or the small number density of galaxies selected in the entire field of view.

In summary, the observed data in the three regions taken together are appear to favor the \gmodel, but the \tmodel\ cannot be completely ruled out due to the large observation errors and the scatters in the model predictions.

\begin{figure}
    \centering
    \includegraphics[width=\linewidth]{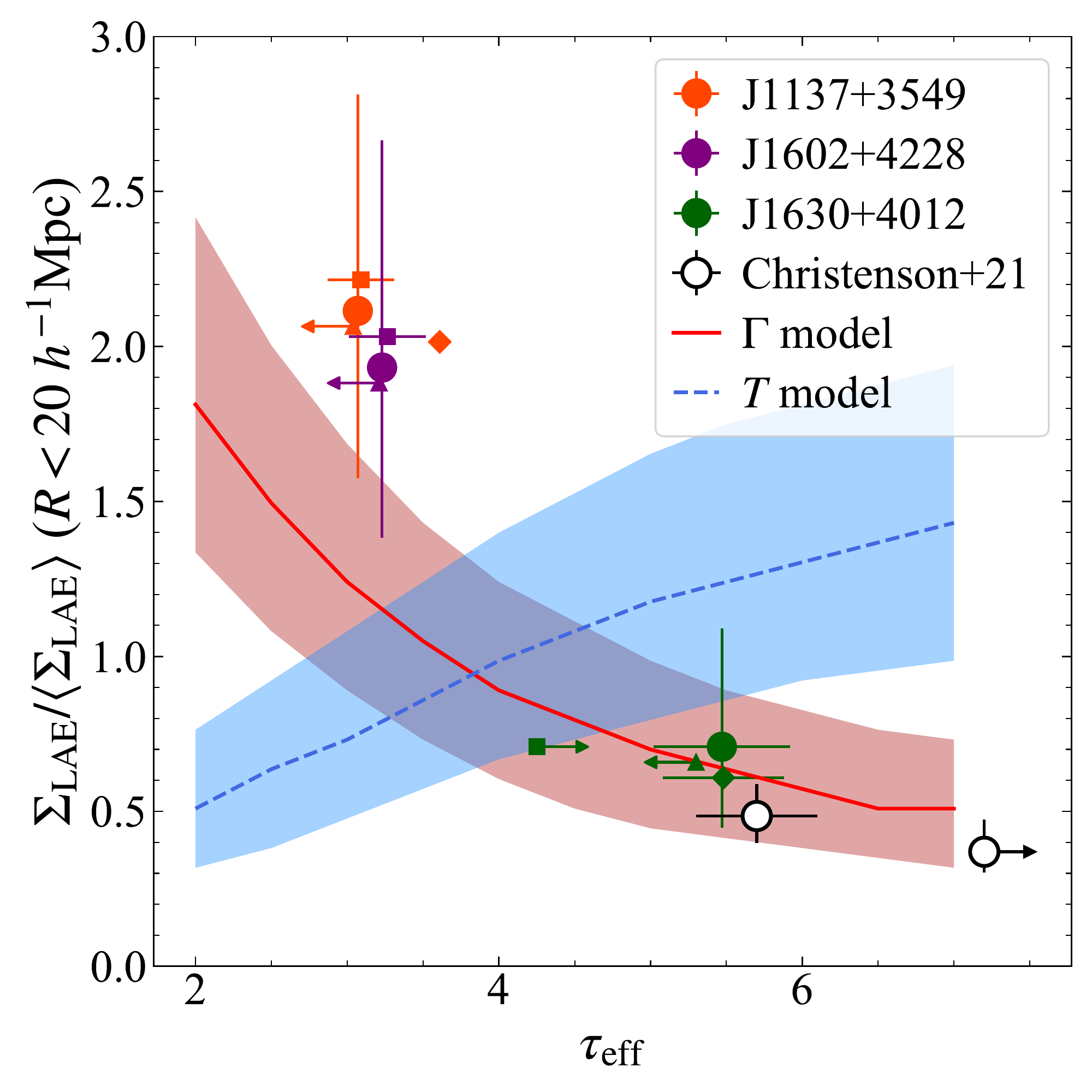}
    \caption{The relation between $\taueff$ and LAE density within 20 $h^{-1}$Mpc from the quasar sightline.
    The red and blue lines indicate the model predictions of the \gmodel\ and the \tmodel\ with the 68\% ranges from random trials shown by the shaded regions as in Figure \ref{fig:radial_los}.
    The filled circles show the $\taueff$ measurements for \qsoa\ (orange), \qsob\ (purple), and \qsoc\ (green), respectively. 
    The open circles indicate the measurements of \citet{Christenson2021}.
    The triangles, diamonds, and squares show the $\taueff$ upper limit using peak flux, $\taueff$ from photometry, and $\taueff$ when changing the zero point by $\pm 0.1$ mag, respectively, and are slightly shifted vertically for clarity.
    The vertical error bars present 1$\sigma$ Poisson errors.
    }
    \label{fig:tau_sigma}
\end{figure}

Figure \ref{fig:tau_sigma} shows the relation between $\taueff$ at the quasar sightline and the LAE density around it within 20 $h^{-1}$Mpc.
The points represent the $\taueff$ estimated by several ways described in Sec \ref{sec:check_tau}.
Although there are variations in the measured $\taueff$, the LAE density in the \qsoc\ field is suggestive of the \gmodel.
In the case of \qsoa\ and \qsob, the observed LAE density is larger than either model, but more consistent with the \gmodel; the \tmodel\ prediction is well below the observed point. 
It should be noted that with this degree of high and low-$\taueff$ of our sample, it is somewhat difficult to distinguish the two models. 
Together with previous observations \citep{Christenson2021} for other two high-$\taueff$ regions, we conclude that the overall observational results are well consistent with the \gmodel.

One of the major factors giving ambiguity to the conclusion is the $\taueff$ uncertainty. 
As can be seen in Figure \ref{fig:qspec}, the Gunn-Peterson troughs of our sample are not long enough to cover the whole transmission wavelengths of the NB816 filter, unlike that of J0148+0600, which has an exceptionally long trough.
The $\taueff$ uncertainty is small, if the trough is sufficiently long, as in J0148+0600, whereas, it is difficult to measure the appropriate $\taueff$ in three regions targeted by this study where there are fine structures of the IGM transmission in the NB816 wavelength range.
When we measure the $\taueff$ of \qsoa\ and \qsob\ considering the exact NB816 transmission curve, their $\taueff$ turn out to be as high as $\sim5$, and consequently, it could alter our statistical conclusions in comparison with the models.
It is hard to come to a clear conclusion until we measure the redshifts of galaxies around the sightlines.
On the other hand, $\taueff$ fluctuations due to unstable IGM transmission in the filter wavelength range also dilute the $\taueff$-galaxy density relation in the model predictions.
In the sense, the $\taueff$ value predicted by the model
to be compared with should be calculated taking into account
the NB filter transmission, which would significantly affect the observed redshift distribution of LAEs.
Nevertheless, we would like to emphasize that, in Figure \ref{fig:radial_los} and \ref{fig:tau_sigma}, the $\taueff$ for both the observation and the model are measured in the 50 $h^{-1}$Mpc range, making them an appropriate comparison. 

To avoid the influence of this $\taueff$ uncertainty, it is ideal to select almost perfect opaque/transparent regions over $>$50 $h^{-1}$Mpc.
Several quasar sightlines with significantly long ($>$80 $h^{-1}$Mpc) troughs have been found recently \citep{Zhu2021}. 
However, the number of such sightlines is still small, and even smaller if we further restrict the number to those that fit the NB wavelength range; therefore, it is challenging to increase the size of ideal sample.
Another way to overcome the problem is to make more detailed model predictions according to the small scale $\taueff$ fluctuations, which are involved in the large model uncertainties, though as discussed later, even this is taken into account, cosmic variance or LAE bias can make it difficult to distinguish the two models.
Third, spectroscopic observation of all LAEs around sightlines will clearly reveal the relation between LAE density and IGM transmission in sightline direction.
Spectroscopic follow-up of all the LAEs is very expensive now, but will become much easier in future with next-generation large multi-object spectrographs such as the Prime Focus Spectrograph (PFS) on the Subaru Telescope or the Multi-Object Optical and Near-infrared Spectrograph (MOONS) on the Very Large Telescope (VLT).

It is interesting to compare with a observational result by \citet{Meyer2020}, who detected $2-3\sigma$ excess of \lya transmission spikes on the sightlines of background quasars on scales 10–60 cMpc around spectroscopically confirmed galaxies; however, they also found that some large transmission spikes are not associated with any detected galaxy. 
Both $\Gamma$ and $T$ may contribute to the fluctuation of $\taueff$, and perhaps the contribution varies from place to place,
so that even in similarly high-$\taueff$ region, the galaxy densities could be different.
The question is what determines whether $\Gamma$ or $T$ is dominant for a given place, but that is not clear from this study alone.
The process of reionization may be more complicated than we thought, if fluctuations in $\Gamma$ and $T$ and neutral islands all come into play. 
The overdensities of J1137+3549 and J1602+4228 near the sightlines are much higher than expected for their $\taueff$ (Figure \ref{fig:tau_sigma}),
so factors other than the reionization process may be at play.
It is increasingly important to examine multiple fields like this study.

\subsubsection{Late ionization model}
Like the \gmodel, the late ionization model also predicts galaxy deficit in high $\taueff$ regions \citep{Nasir2020,Keating2020},
which is consistent with our observational result of J1630+4012.
It is difficult to distinguish the late reionization model, whose mean LAE surface density profile for opaque sightlines
should generally show underdensity, from the \gmodel.
Interestingly, \citet{Keating2020} suggested that, in the late reionization model, the underdense regions corresponds to high-$\taueff$ regions only before the neutral islands have been ionized, and after that these hot, recently reionized voids should instead correspond to the low-$\taueff$ regions.
This is in clear contrast with the \gmodel, whose low-$\taueff$ sightlines almost always correspond to overdensities.
Our observational results for low-$\taueff$ regions in the \qsoa\ and \qsob\ fields show apparent overdensities around the sightlines, which is naively consistent with the \gmodel, and likely to be inconsistent with the late reionization model.
However, \citet{Nasir2020} suggested that some of the most transmissive regions in the late reionization model should actually correspond to the deficit of LAEs, but they are relatively rare, and something similar could be observed even in the \gmodel.
Based on their model, both the \gmodel\ and the late reionization model predict galaxy densities with large scatter in the low-$\taueff$ region, so it is difficult to distinguish between the two models, even in the low-$\taueff$ region.
The LAE radial distribution based their model including the reionization temperature fluctuation shows a larger scatter than \citet{Davies2018}.
Our result, which shows relatively strong correlation between low-$\taueff$ and LAE overdensity, may at least suggest that the relic temperature fluctuation does not affect reionization that much.
However, the number of our low-$\taueff$ sample is limited to only two, and most of the simulations only care about the most opaque and transmissive sightlines, which does not allow for a rigorous comparison with this study.
More quantitative comparisons are required in the future.

\subsubsection{Quasar model}
To account for the observed dispersion of $\taueff$ at $z > 5.5$, there is an alternative model, in which rare sources like
quasars or AGNs could generate substantial large-scale ($\sim50\ h^{-1}$ Mpc) opacity variations \citep{Chardin2017}. 
If quasars exist in these fields, quasars with extremely high \lya EWs should be detectable in our NB imaging observation. 
There are two such sources in the \qsoa\ field only as very bright sources with $\nb<23$, which could be quasar candidates.
We check the existing database and they do not seem to be already spectroscopically confirmed sources. 
Both of them are at the edge of the field of view, one in the LAE high-density region and the other in the low-density region. 
Whether these are quasars or not will not be known until spectroscopy is taken.
However, the strong correlation between the $\taueff$ and galaxy density, consistently seen in all the observed two low-$\taueff$ and three high-$\taueff$ sightlines (i.e., this work plus \citet{Becker2018}, \citet{Kashino2020}, and \citet{Christenson2021}) is rather likely to be contradict this quasar model that predicts the lack of clear correlations between them.
In addition, their model assumes a high number density of quasars with $\sim10^{-6}$ Mpc$^{-3}$ based on \citet{Giallongo2015},
which is in contradict with some measurements \citep{Kashikawa2015,Onoue2017,Matsuoka2019,Jiang2022}, and is also in
tension with constraints on the \ion{He}{ii} reionization \citep{Worseck2014}.

\subsection{LAE large-scale spatial bias}\label{sec:laebias}
As seen in Figure \ref{fig:tau_sigma}, the overdensities in the vicinity of the sightlines in the \qsoa\ and \qsob\ fields are higher than the $\Gamma$ model prediction.
This indicates that there may be factors other than the reionization process.
We have detected LAEs at $z= 5.726\pm0.046$, where the \lya emission lines are easily detected by the NB816 imaging observations  under the assumption that LAEs are representative galaxies at the epoch.
However, if there is some bias in the spatial distribution of LAEs and they do not represent the average galaxy distribution in the universe, we cannot expect to see the relationship between \lya opacity and LAE local density as expected from the theoretical models.
Some studies indicated physical similarities between LAEs and non-LAEs \citep{Hathi2016,Shimakawa2017}, suggesting that LAEs can be used to probe the general low mass star-forming galaxies tracing the underlying
density structures, while several recent studies have pointed out that the distribution of LAE is different from the field, especially in high density regions.
Recently, \citet{Ito2021} claimed that the cross-correlation signals between LAEs and star-forming galaxies are significantly lower
than their auto-correlation signals up to $\sim30$ cMpc, suggesting that the distribution of LAEs are different from those of general galaxy populations. 
\citet{Toshikawa2016} showed that the \lya EW is systematically low in high-density regions of LBG, suggesting possible \lya suppression in galaxy overdense regions.
\citet{Shimakawa2017} have found a LAE number deficit in a protocluster core composed of H$\alpha$ emitters at $z = 2$  on scales of $\sim$ a few Mpc.
\citet{Shi2019} also found a spatial offset of $\sim$ a few tens of Mpc between the density peak of LAEs and LBGs, suggesting their different age or different dynamic stages.
\citet{Cai2017}, \citet{Momose2020}, and \citet{Liang2021} showed LAEs tend to avoid the highest \ion{H}{i} density regions.
More recently, \citet{Huang2022}  conducted a LAE search around the \textit{Hyperion} protocluster at $z\sim2.47$, and found that LAE well traces the large scale structures. However, their result also suggested that \lya emission is suppressed in the highest \ion{H}{i} regions.
These results indicate that the LAE may not be a good tracer of underlying large-scale structures.
If this hypothesis is correct, it is difficult to distinguish between the \gmodel\ and the \tmodel\ using LAEs as in this study.
Independent validation using another galaxy population, e.g. LBG \citep{Kashino2020}, is required.

\section{summary}
\label{sec:summary}

In this study, we conduct HSC imaging with the NB816 filter to perform a LAE search at $z\sim5.7$ in two fields containing background quasars, whose sightlines show low \lya optical depth ($\taueff\sim3$) at the same redshift, and a field with high optical depth ($\taueff\sim5.5$). 
Our goal is to test two conflicting models for the origin of large scatter of \lya optical depth at $z>5.5$.
In the \gmodel, the observed large $\taueff$ fluctuation is due to the fluctuation in the galaxy-dominated UV background, and low(high) galaxy density should generally be observed at high(low) $\taueff$ regions.
In the \tmodel, $\taueff$ fluctuation is due to the fluctuation in the IGM gas temperature, and high(low) galaxy density should generally be observed at high(low) $\taueff$ regions, contrary to the \gmodel.

The major results are summarized below.

\begin{enumerate}
\item We estimate $\taueff$ in 50 $h^{-1}$Mpc range, using PCA continuum estimation.
The resultant $\taueff$ values are $3.07\pm0.03$, $3.23\pm0.05$, and $5.47\pm0.86$ for \qsoa, \qsob, and \qsoc, respectively.
\qsoa\ and \qsob\ show low $\taueff$, which are in <17\% from the bottom in the $\taueff$ distribution at $z=5.6-5.8$ \citep{Bosman2022}, and this work is the first to investigate galaxy density in such low $\taueff$ regions.
In contrast, \qsoc\ shows high $\taueff$, which exceeds the 95\% range in $\taueff$ distribution  predicted by the uniform UV background model \citep{Becker2015} and cannot be explained by the density variations alone.
We evaluate the uncertainties of the $\taueff$ measurement in several ways. 
Each measurement agrees almost consistently
with the others: the $\taueff$ of J1137+3549 and J1602+4228 are $\sim$3,
while the $\taueff$ of J1630+4012 is $\sim$5.5, although there are non-negligible uncertainties.
\item 
The numbers of LAE candidates are 84, 80, and 97 for the field of \qsoa, \qsob, and \qsoc, and the survey area is 5094, 5025, and 4826 arcmin$^2$, respectively. 
We map the spatial distributions of LAE candidates in the three fields by carefully correcting the variation of sensitivity in the field of view using mock LAEs.
\item 
LAE overdensities are found within 20 $h^{-1}$Mpc of the quasar sightlines in the \qsoa\ and \qsob\ fields, while an LAE underdensity is found in the \qsoc\ field.
These over(under)densities have the significance of 2.3$\sigma$, 2.7$\sigma$, and 1.3$\sigma$, respectively.
The radial distributions of LAEs in the \qsoa\ and \qsob\  fields have upward trends toward the quasar sightline, while that of the \qsoc\ field decreases toward the center.

\item
We quantitatively compare the observed spatial distributions of LAEs to the predictions for the $\Gamma$ model and the $T$ model of \citet{Davies2018}. 
The radial distribution of LAE surface number density centered on the quasar sightline of \qsoc\  is consistent with the \gmodel, while the profiles for the other two fields are rather consistent with the $\Gamma$ model, though the $\taueff$ values in these fields are not extreme enough to distinguish between the models, and cannot determine which model is more plausible.
\item
We, therefore, use a more robust statistic, the \lya opacity-galaxy density relation (Figure \ref{fig:tau_sigma}): the relation between IGM $\taueff$ at the quasar sightline and the LAE density around it within 20 $h^{-1}$Mpc. 
The results of all the three fields along with the previous observations are found to be consistent with the \gmodel.
\item
In the low $\taueff$ regions, in which relatively
strong correlations between low-$\taueff$ and LAE overdensity are observed, which may suggest that the relic temperature fluctuation does not affect reionization that much.
Another possibility is that LAE is not a good tracer of underlying large-scale structures and thus affects galaxy density somewhat independently from reionization.

\end{enumerate}

The concern that remains in relation to the last point above is  that LAE could not be used as a representative of galaxies.
Although \citet{Becker2018} and \citet{Kashino2020} showed consistent results using LAE and LBG, we can not deny that LAE shows different distribution from other galaxy populations, or absorption by neutral \ion{H}{i} changes the apparent distribution of LAEs.
The search for continuum-selected galaxies, such as LBGs, in the same fields is required to confirm our findings focusing on LAEs.
In some of our fields, the LBG search has also been conducted based on the similar motivation.
In the future, we will compare the results of LAE and LBG in the same field and discuss the consistency between different galaxy populations.

In addition, observation of much lower-$\taueff$ regions is also needed.
This study is the first to investigate the LAE density at low-$\taueff$ regions, but the $\taueff$ values are not low enough to clearly distinguish between the \gmodel\ and the \tmodel.
Observations of low-$\taueff$ regions will give further insight into the plausible model for the physical origin of the patchy reionization.
Also in the high-$\taueff$ region, only three regions have been observed, and we need to increase the number of sample to find out what is causing the difference between the fields.
To obtain more conclusive results, we need to improve the model and further observations.

\section*{Acknowledgements}
We appreciate the referee, George Becker, for his helpful suggestions and comments, which significantly improves the paper.
We are grateful to Frederick Davies for providing the model predictions recalculated to fit our observations.
We thank the \textit{hscpipe} helpdesk for many helpful suggestions.
We also thank to Satoshi Yamanaka, who kindly provided the code to measure the limiting magnitude.
RI acknowledges support from JST SPRING, Grant Number JPMJSP2108.
This research was supported by the Japan Society for the Promotion of Science through Grant-in-Aid for Scientific Research 21H04490.

This research is based on data collected at Subaru Telescope,
which is operated by the National Astronomical Observatory of Japan.
We are honored and grateful for the opportunity of observing the Universe from Maunakea, which has the cultural, historical and natural significance in Hawaii.

This work has made use of data from the European Space Agency (ESA) mission
{\it Gaia} (\url{https://www.cosmos.esa.int/gaia}), processed by the {\it Gaia}
Data Processing and Analysis Consortium (DPAC,
\url{https://www.cosmos.esa.int/web/gaia/dpac/consortium}). Funding for the DPAC
has been provided by national institutions, in particular the institutions
participating in the {\it Gaia} Multilateral Agreement.

\section*{Data Availability}
The data underlying this article will be shared on reasonable request to the corresponding author.




\bibliographystyle{mnras}
\bibliography{bibtex} 




\appendix
\section{Surface density of LAEs down to a fixed limiting magnitude}\label{sec:appendix}

\begin{figure}
    \centering
    \includegraphics[width=\linewidth]{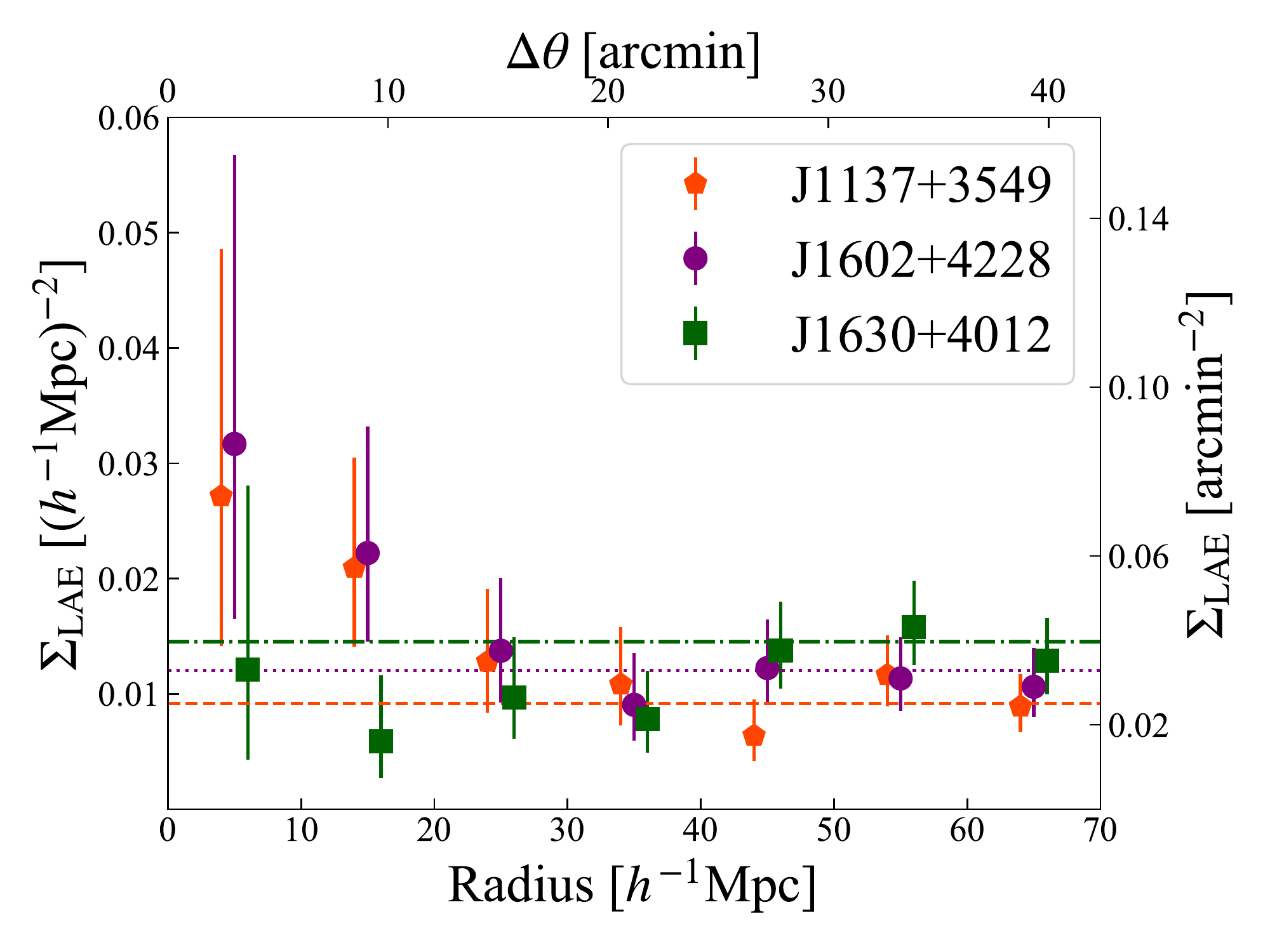}
    \caption{Same as Figure \ref{fig:radial_density}, but the limiting magnitude are fixed to 25.2 mag in all fields.
    The data points are slightly shifted horizontally for clarity.}
    \label{fig:radial_shallow}
\end{figure}

Figure \ref{fig:radial_shallow} shows the surface density of LAEs as a function of projected distance from the quasar sightline down to the fixed limiting magnitude of 25.2 mag, which is the shallowest among the three fields. 
The profiles show the similar trend to that in Figure \ref{fig:radial_density}, although the profiles of \qsoa\ and \qsoc\ are shifted slightly lower.

\bsp	
\label{lastpage}
\end{document}